\documentclass[sigconf]{acmart}

\usepackage{array}
\usepackage{makecell}
\usepackage{multirow}
\usepackage{lscape}
\usepackage{graphicx}
\usepackage{geometry}
\usepackage{colortbl}
\usepackage{lipsum}
\usepackage{listings}
\usepackage{enumitem}

\AtBeginDocument{%
  \providecommand\BibTeX{{%
    \normalfont B\kern-0.5em{\scshape i\kern-0.25em b}\kern-0.8em\TeX}}}

\copyrightyear{2024}
\acmYear{2024}
\setcopyright{rightsretained}
\acmConference[DIS '24]{Designing Interactive Systems Conference}{July 1--5,
2024}{IT University of Copenhagen, Denmark}
\acmBooktitle{Designing Interactive Systems Conference (DIS '24), July 1--5
2024, IT University of Copenhagen,
Denmark}
\acmDOI{10.1145/3643834.3661547}
\acmISBN{979-8-4007-0583-0/24/07}

\acmJournal{JACM}
\acmVolume{37}
\acmNumber{4}\acmArticle{111}
\acmMonth{8}

\acmPrice{15.00}
\acmISBN{978-1-4503-XXXX-X/18/06}

\begin{document}


\definecolor{myRed}{RGB}{255, 45, 85}
\definecolor{myGreen}{RGB}{76, 217, 100}
\definecolor{myBlue}{RGB}{0, 122, 255}
\definecolor{myOrange}{RGB}{255, 136, 0}
\definecolor{darkergreen}{RGB}{37, 155, 57}
\definecolor{myTeal}{RGB}{181, 255, 251}
\definecolor{myLightGray}{RGB}{213, 217, 224}
\newcommand{\red}[1]{{\color{myRed}{#1}}}
\newcommand{\green}[1]{{\color{myGreen}{#1}}}
\newcommand{\blue}[1]{{\color{myBlue}{#1}}}
\newcommand{\orange}[1]{{\color{myOrange}{#1}}}
\newcommand{\darkgreen}[1]{{\color{darkergreen}{#1}}}
\newcommand{\teal}[1]{{\color{myTeal}{#1}}}
\newcommand{\lightGray}[1]{{\color{myLightGray}{#1}}}

\newcommand{\jl}[1]{\orange{JL: #1}}

\title{How People Prompt Generative AI to Create Interactive VR Scenes}

\begin{abstract}

Generative AI tools can provide people with the ability to create virtual environments and scenes with natural language prompts.
Yet, \textit{how} people will formulate such prompts is unclear---particularly when they inhabit the environment that they are designing.
For instance, it is likely that a person might say, ``Put a chair here,'' while pointing at a location.
If such linguistic and embodied features are common to people's prompts, we need to tune models to accommodate them.
In this work, we present a Wizard of Oz elicitation study with 22 participants, where we studied people's implicit expectations when verbally prompting such programming agents to create interactive VR scenes.
Our findings show when people prompted the agent, they had several implicit expectations of these agents: (1) they should have an embodied knowledge of the environment; (2) they should understand embodied prompts by users; (3) they should recall previous states of the scene and the conversation, and that (4) they should have a commonsense understanding of objects in the scene.
Further, we found that participants prompted differently when they were prompting \textit{in situ} (i.e. within the VR environment) versus \textit{ex situ} (i.e. viewing the VR environment from the outside).
To explore how these lessons could be applied, we designed and built Ostaad, a conversational programming agent that allows non-programmers to design interactive VR experiences that they inhabit.
Based on these explorations, we outline new opportunities and challenges for conversational programming agents that create VR environments.
\end{abstract}

\author{Setareh Aghel Manesh}
\affiliation{%
  \institution{University of Calgary}
  \streetaddress{2500 University Drive NW}
  \city{Calgary}
  \state{Alberta}
  \country{Canada}
  \postcode{T2N 1N4}
}
\email{saghelma@ucalgary.ca}

\author{Tianyi Zhang}
\affiliation{%
  \institution{Singapore Management University}
  \streetaddress{80 Stamford Rd}
  \city{Singapore}
  \country{Singapore}}
\email{tianyizhang.2023@phdcs.smu.edu.sg}

\author{Yuki Onishi}
\affiliation{%
  \institution{Singapore Management University}
  \streetaddress{80 Stamford Rd}
  \city{Singapore}
  \country{Singapore}}
\email{yukionishi@smu.edu.sg}

\author{Kotaro Hara}
\affiliation{%
  \institution{Singapore Management University}
  \streetaddress{80 Stamford Rd}
  \city{Singapore}
  \country{Singapore}}
\email{kotarohara@smu.edu.sg}

\author{Scott Bateman}
\affiliation{%
  \institution{University of New Brunswick}
  \streetaddress{Sir Howard Douglas Hall}
  \city{Fredericton}
  \state{New Brunswick}
  \country{Canada}
  \postcode{E3B 5A3}
  }
\email{scottb@unb.ca}

\author{Jiannan Li}
\affiliation{%
  \institution{Singapore Management University}
  \streetaddress{80 Stamford Rd}
  \city{Singapore}
  \country{Singapore}}
\email{jiannanli@smu.edu.sg}

\author{Anthony Tang}
\affiliation{%
  \institution{Singapore Management University}
  \streetaddress{80 Stamford Rd}
  \city{Singapore}
  \country{Singapore}}
\email{tonyt@smu.edu.sg}

\renewcommand{\shortauthors}{Aghel Manesh and Zhang, et al.}





\begin{teaserfigure}
    \includegraphics[width=\textwidth]{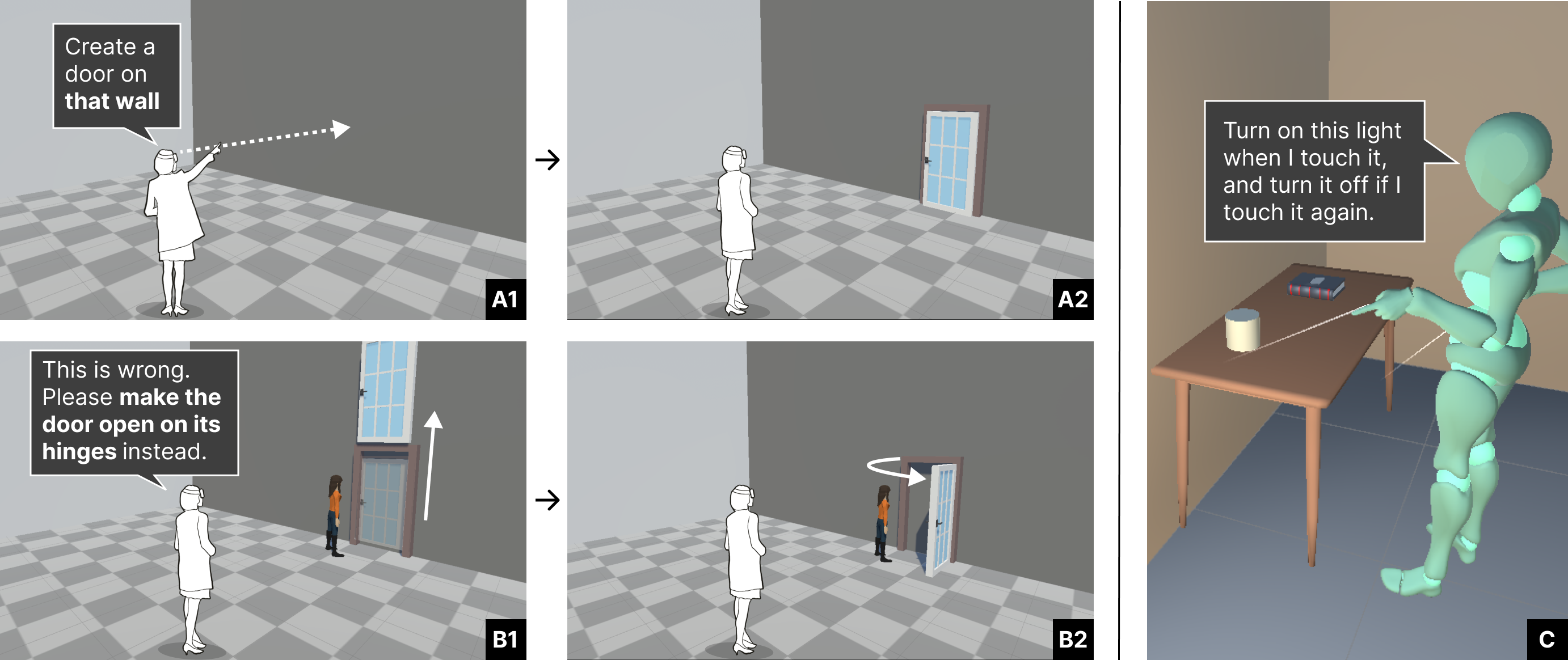}
    \caption{Through an elicitation study, we identified four implicit expectations people have when they prompt an intelligent agent to create virtual environments. As illustrated in (A1, A2), people assume agents have \textbf{embodied knowledge}---an awareness of the objects in the environment, and their use of \textbf{embodied gestures} to communicate their intentions are expected to be understood. As illustrated in (B1, B2), people assume agents have \textbf{common sense knowledge} about how artifacts (e.g. doors and hinges) should function. (C) Based on these findings, we built Ostaad, a programming agent for creating interactive virtual environments.}
    \label{fig:teaser}
    \Description{Five pictures show the user exists in the study VR environment. The first two pictures illustrate that the user prompts to create a door object. The user uses a pointing gesture to indicate the door location while she's prompting in the first figure, and then the door shows up in the next figure. The next two pictures illustrate that the user prompts to attach the door function. Although the user expected to agent that the door would be opened the common way, the door opened the wrong way in the third figure. After that, she modified the prompt, and the door opened the expected way. The right biggest figure illustrates the virtual scene of our developed Ostaad system, the user can point to the object and prompt their function. }
\end{teaserfigure}

\maketitle

\section{Introduction}

Generative AI tools enable the exploration of creative outputs using natural language prompting, allowing for rapid creation and iteration of multimedia content (e.g.,~\cite{epstein_art_2023,zhou_generative_2023}).
The code generation ability of current LLMs further powers programming agents that synthesize computer programs supporting complex logic and user interactions from language input (e.g.,~\cite{svyatkovskiy_intellicode_2020,zhang_demystifying_2023}). 
An exciting opportunity for generative tools involves assisting in the design of interactive 3D virtual environments with natural language prompts.
For instance, recent work \cite{de_la_torre_llmr_2023} demonstrates the use of general-purpose LLMs and diffusion models to create 3D scenes, 3D objects and environments based on a simple set of primitives. 
Appropriately designed programming agents that respond to natural language descriptions could enable a whole class of non-specialists to create rich, interactive environments responsive to inhabitants' behaviours and needs.
These environments could serve as prototypes for informing the development of intelligent virtual worlds, such as training simulations and video games, or physical worlds, such as smart homes and factories. 
Our work ultimately aims to democratize the creation and design of such interactive environments, including the creation and refinement of interactive behaviours---typically the domain of specialist programmers.

Yet, the problem is that we do not understand \textit{how} people expect to be able to prompt, create, and refine such interactive scenes and environments.
Without such knowledge, we run the risk of building AI-driven tools with features and capabilities that do not match user expectations. 
This mismatch has been considered one of the major reasons behind human-AI collaboration breakdown~\cite{grimes_mental_2021,kocielnik_will_2019}. 
While prior research has uncovered patterns in end-user behaviors when working with intelligent agents~\cite{zamfirescu-pereira_why_2023,crisan_eliciting_2023,sarkar_what_2022}, these explorations have primarily focused on text-based or 2D tasks on a desktop, such as conversing with a chatbot, or coding.
In contrast, design tasks in 3D environments often entail a distinct set of user behaviors from their 2D counterparts due to the additional spatial affordances.
For example, research on 3D immersive learning and analytics has highlighted people's tendencies and the associated benefits of combining language commands with embodied actions, including physical navigation and gestures~\cite{johnson-glenberg_immersive_2018}.
Similarly, when communicating with other humans in immersive environments, people rely on view frustrums, gesturing and awareness of others' locations to articulate ideas (e.g., ~\cite{hindmarsh_fragmented_1998,fraser_supporting_1999,hindmarsh_object-focused_2000,tang_verbal_2012,kumar_tourgether360_2022}).
Yet, how these may be expressed in verbal prompts to an intelligent agent is unclear.
We are thus interested in understanding users' verbal and nonverbal behaviors when they modify 3D environments through programming agents.

To gather information about how people expect to prompt a programming agent to design an interactive virtual environment, we conducted a Wizard of Oz elicitation study with 22 participants.
Elicitation studies have commonly been used to understand the critical aspects of people's mental models and what interactions are essential to be sensed and understood by the system (e.g., for gestures~\cite{wobbrock_user-defined_2009}).
Our use of an elicitation study here is aligned with this---to understand the linguistic and embodied characteristics of how people would prompt an intelligent agent for design.
We showed participants the intended changes they should make in the scene, such as modifying furniture layouts and programming a proximity-aware lamp, and asked them to prompt a programming agent, who would (via the Wizard of Oz protocol) initiate these changes.
Our goal was to understand what capabilities people implicitly expect these programming agents to have.

\begin{figure*}[th]
  \centering
  \includegraphics[width=\linewidth]{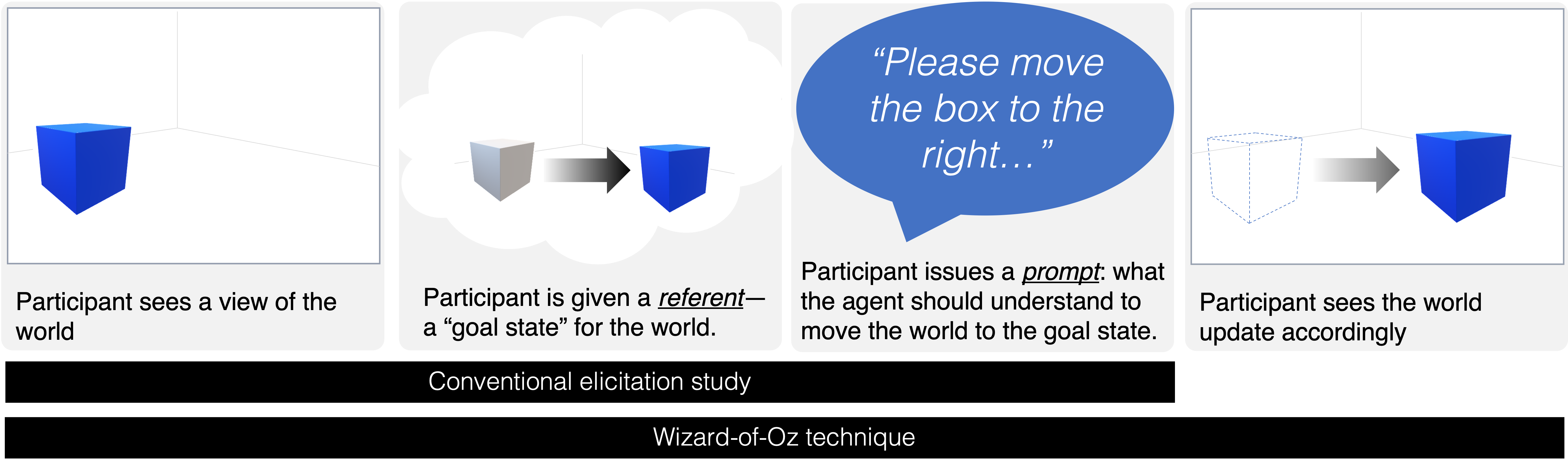}
  \caption{Our elicitation study uses a wizard-of-oz approach to understand how people prompt a programming agent.}
  \label{fig:elicitation+woz}
  \Description{Four figures are explaining our elicitation study scenarios. In the first figure, there is one blue cube object in the virtual world. The second figure illustrates the referent scene transition to the next step, and then the participant issues a prompt to make the referent happen in the third figure. Until here shows the details of the elicitation study. The last figure illustrates the virtual world updated following the employed prompt, and the blue cube moves to the right. It seems automatic, but the reality is that the experimenter controls and updates the virtual world behind a Wizard of Oz method.}
\end{figure*}

As illustrated in Figure~\ref{fig:elicitation+woz}, our participants would see the VR scene and be given a \underline{referent}, which is a goal state for the scene. Their task was to construct a \underline{prompt} that they imagined a programming agent could use to move the scene to the goal. We used 43 such referents, where participants would create and modify scene objects, and create interactive behaviours (e.g., ``Turn on the light when someone points at it.''). Our elicitation focused on understanding how people would construct their prompts, which would provide critical new information about how conversational programming agents should be designed.

Our findings suggest that designing a functional conversational programming agent requires supplying the agent with far more than simply the verbal utterances of the user.
Based on our analysis, people have several implicit expectations of these programming agents:
\begin{enumerate}
    \item First, people expect the agent to have \textbf{embodied knowledge} of the scene---the objects in the scene, how they are arranged within it and in relation to one another.
    \item Second, people expect the agent to understand \textbf{embodied prompts} when they are prompting \textit{in situ}, where they can pair verbal utterances with both explicit gestures (e.g., pointing) as well as implicit references (e.g., where one is looking).
    \item Third, people expect the agent to have \textbf{conversational memory}, where beyond referring to past interaction with the user, can also understand and recall past states of the scene in relation to that interaction.
    \item Fourth, people expect the agent to have \textbf{common sense knowledge} of how objects in scenes should behave.
\end{enumerate}

Based on these findings, we designed Ostaad, a conversational agent which allows people to use embodied voice prompts to create 3D scenes and interactions. 
We built Ostaad to demonstrate the feasibility of applying some of the design implications from the elicitation study to a functional system.
Like previous work \cite{de_la_torre_llmr_2023}, Ostaad is a programming environment that allows people to declaratively prompt an agent with their intentions for what should be in a VR environment and how it should behave. Ostaad departs from previous work \cite{de_la_torre_llmr_2023} by allowing users to prompt the system verbally while \textit{in situ} in the VR environment. As such, Ostaad begins to accommodate proxemic variables related to the user's location and visual field of view, as well as their bodily gestures (e.g., pointing) in relation to scene objects. 

The main contribution of our work is an articulation of how \textit{embodied programming agents} of the future should be designed. To illustrate how this can be done, we take initial steps toward realizing one such an agent. Beyond the technical/programmatic capabilities of such agents, we articulate a set of requirements for their conversational/awareness ability with users. Critically, programming agents should understand the state of the world, the user's relationship with that world, and be able to interact with the user to support the user's design vision.

\section{Related Work}

To situate our work, we briefly describe recent work on easing the challenge of designing 3D virtual environments. In relation to our interest, we then discuss lessons from human-AI collaboration explorations, and how these relate to how agents have been used to support programming agents. Finally, to ground our methodological approach, we briefly explain prior use of elicitation studies and the Wizard of Oz technique.

\subsection{Authoring 3D Virtual Environments}

Traditionally, 3D virtual environments have been created using desktop software for 3D modelling and animation (for creating objects and their movements), and game engines (to compose environments and add interactivity). 
However, these tools have steep learning curves.
To lower the barrier for more creators, researchers have explored \emph{in situ} authoring, which immerses users in the environments being edited through virtual reality display and input technologies.
The in situ paradigm allows users to directly work within the 3D space to be built, creating and modifying 3D objects through intuitive embodied input such as direct manipulation with hands or sketching~\cite{arora_symbiosissketch_2018,leiva_pronto_2020}.
Some work further supported generating 3D models by scanning real-world objects or models~\cite{wang_gesturar_2021,sra_oasis_2018}.
Research has also explored how object modelling can be extended to support authoring interactive behaviors.
For example, users may employ direct manipulation to set object poses in a series of keyframes to animate an object~\cite{arora_magicalhands_2019}.
Tools for creating animations could additionally record the desired input for initiating certain reactions from the environments (e.g., turning on a light when a user approaches it), allowing users to program such interactive behaviors by demonstration~\cite{leiva_pronto_2020}.
Another approach employs visual block-based programming, which allows more complex interaction at the expensive visual complexity~\cite{zhang_flowmatic_2020,ens_ivy_2017}.

In comparison to the aforementioned approaches, language-based authoring methods have a unique advantage.
They allow users to flexibly express high-level intentions without requiring significant training.
While early work already proposed language-based 3D model retrieval and scene composition~\cite{seversky_real-time_2006,ma_language-driven_2018}, recent breakthroughs in large multi-modal generative models have significantly expanded the range of vocabularies and intents such methods can understand and execute (e.g.,~\cite{de_la_torre_llmr_2023}).
Our work aims to reveal users' expected interactions with these emerging capabilities.

\subsection{Human-AI Communication}
As generative AI, especially LLMs, show strong promise in empowering individual creators, the research community is increasingly interested in understanding whether and how non-expert users could effectively prompt these tools to achieve desired outcomes~\cite{zamfirescu-pereira_why_2023,dang_how_2022}.
Studies found that people often struggle with crafting effective prompts. 
In addition to common end-user programming challenges ~\cite{ko_six_2004}, people also overly attribute human-like capabilities to AI models, e.g., expecting them to resolve ambiguous instructions~\cite{zamfirescu-pereira_why_2023}.
This mismatch between user expectations and AI model capabilities speaks to the emergence of the intricate art of prompt engineering and highlights the difficulty for non-expert users to work with LLMs using the native text interface (e.g., OpenAI Playground). 
To overcome this conceptual barrier, a number of tools aimed to scaffold human-AI communication with known-to-be-successful practices~\cite{chung_talebrush_2022,feng_canvil_2024,wu_promptchainer_2022}, such as: breaking large tasks into steps~\cite{wu_ai_2022}, automating the creation of suggestions for good prompts~\cite{brade_promptify_2023}, and exposing the key features of expert prompts through graphical interfaces~\cite{kim_cells_2023}.

While effective, prior study designs have generally assumed a desktop and 2D graphical user interface setting.
Our research aims to provide insights into user expectations when verbally prompting AI models in an embodied context---both in terms of how they physically gesture, and in terms of the referential resources they can use.
This should bridge the expectation-capability gap and inform the design of future tools.

\subsection{Agents for Programming Intelligent Behaviors}
Pre-trained generative models can generate text, code, and images conditioned on language input~\cite{vaswani_attention_2017,podell_sdxl_2023,chen_evaluating_2021}. 
Beyond directly adopting the generated content, people are further interested in leveraging the powerful reasoning abilities~\cite{brown_language_2020} of LLMs to build autonomous agents that can execute complex tasks independently~\cite{yao_react_2022,shinn_reflexion_2023}.
In addition to the ability to interpret user instructions, these agents understand the environments they operate in, remember the history of instructions and their results, and use custom tools to achieve their goals~\cite{wang_survey_2023}.
A common approach for tool usage is to synthesize functional computer programs and run them in the task environment~\cite{gao_pal_2023,liang_code_2023}.
Recent work has explored using LLMs to produce programs for agents to alter 2D and 3D virtual environments~\cite{zhang_proagent_2023,huang_language_2022}, and even the physical world~\cite{huang_inner_2023,ichter_as_2023}, based on users' language instructions.

Our efforts complement existing work that assumed perfect user input and focused on strengthening agent capabilities.
We look at user behaviors with a programming agent for 3D virtual environments, including users' choice of spoken and bodily language and their coping strategies with agents' errors, to inform the design considerations of future agents.

\subsection{Method: Eliciting User Expectations through Wizard of Oz}

Our approach employs both an elicitation study approach and a Wizard of Oz technique (\autoref{fig:elicitation+woz}). 
In a canonical gesture elicitation study~\cite{wobbrock_user-defined_2009}, a participant is shown the starting state of the system and the expected end state of the system (called a \textit{referent}). They are then asked to perform a gesture (a hand posture or movement) that would cause the system's state to move to its end state. This approach informs designers about people's mental models of what aspects of the interaction are important for the system to be able to track, and how the system should sense/detect these interactions. While generally used for gesture design (e.g.,~\cite{villarreal-narvaez_systematic_2020, chan_user_2016}), researchers have also used it to explore how people expect conversational agents to respond (e.g., ~\cite{volkel_eliciting_2021,cambre_methods_2020}), or to explore interaction techniques in multi-display environments (e.g.,~\cite{seyed_eliciting_2012,morris_web_2012}), as well as mixed reality scenarios (e.g.,~\cite{pham_scale_2018}).

Wizard of Oz approaches originated from studies of natural language dialogue systems~\cite{dahlback_wizard_1993}, where researchers were focused on understanding how users would attempt to use such systems. These studies sought to understand how people would react to the system, and how they would attempt to use the system. From the user's perspective, the system appeared to function as described; however, an experimenter controlled the system responses. Such an approach allowed experimenters to ignore system limitations of the time (e.g., speed and accuracy) in favor of understanding the breadth of people's inputs to the system. This approach has been widely adopted in exploring design spaces for human-agent interaction, be it with intelligent agents (e.g.,~\cite{maulsby_prototyping_1993, price_off_2002}), conversational agents (e.g.,~\cite{eberhart_wizard_2022,kuoppamaki_designing_2023}, or for exploring human-robot interactions~\cite{riek_wizard_2012}. 

Our study adopted a hybrid of these approaches. The elicitation technique allowed us to control the space of possible ``expected outputs'' from the system. Participants used the system as if it were functional.

\section{Elicitation Study}

Our goal was to understand how people would prompt an intelligent agent to create objects, modify objects, and to create interactive possibilities in a VR scene. Such intelligent agents would help democratize the design of such spaces, allowing creators to design without needing specialized training. For example, a prompt that a user might utter could be, ``Create a lamp here,'' where the agent would interpret that phrase, and instantiate a lamp object in the VR scene at the appropriate location. Yet, designing such agents requires understanding \textit{how} people expect to be able to prompt such agents---not only in the language people use, but further, how they express themselves, and what they expect the agents to understand.

While our inquiry was exploratory in nature, we designed our study around three research questions:

\begin{enumerate}
    \item RQ1: How would people's prompts vary depending on whether they prompt the agent \textit{in situ} versus \textit{ex situ}?
    \item RQ2: Would people prompt verbally? 
    \item RQ3: How do people respond if the agent creates something unexpectedly?
\end{enumerate}

Our intention was to understand how people constructed prompts and expressed themselves when instructing the intelligent agent for authoring virtual scenes. Identifying and characterizing the linguistic and embodied characteristics of their prompts would help reveal the implicit assumptions people had of the agent's capabilities.

\begin{figure*}[t]
  \centering
  \includegraphics[width=\linewidth]{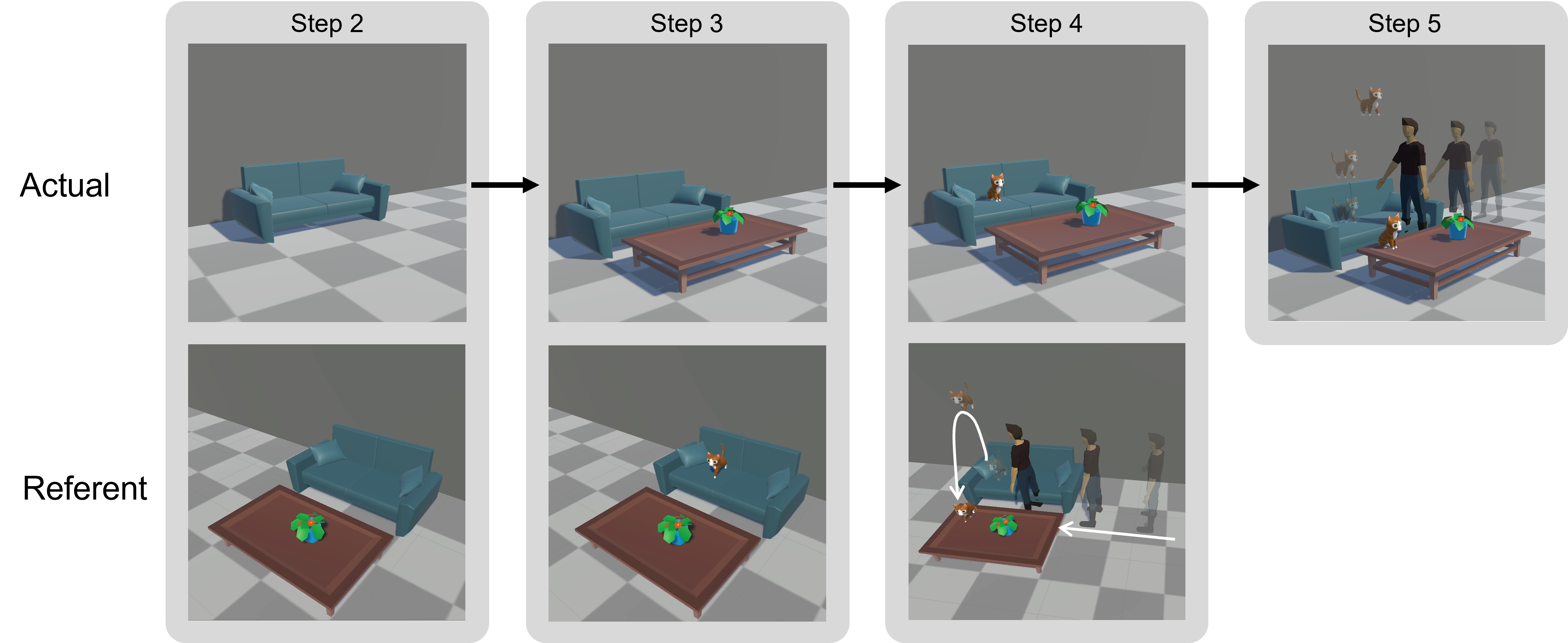}
  \caption{An example of a normal referent from Scene 4 ``Scaredy Cat''. The referents illustrate what the "next state" is. From left to right: (Step 2) Participants see the sofa in the room, and are asked to create a coffee table and a flower pot; (Step 3) Participants now see a sofa, coffee table and flower pot, and are asked to create a cat on the sofa; (Step 4) Participants see the cat, and are asked to create an interaction where the cat jumps to the coffee table if a person gets too close. Step 5 shows the outcome of Step 4. Steps 0 and 1 are left out for brevity.}
  \label{fig:normal-referent}
  \Description{This figure illustrates the example scene from the user study. There are four steps before and after prompting. Upper low figures are happening virtual world view, and lower figures of virtual world views are called referent. Here, firstly there is a sofa, and then the participant asks to create a coffee table with a flower pot and a cat. In step four, the referent shows the interaction where the cat jumps to the coffee table when the person approaches. The next step shows the outcome of step four presenting that the participant prompted successfully.}
\end{figure*}

\subsection{Method}

\begin{figure*}[h]
  \centering
  \includegraphics[width=\linewidth]{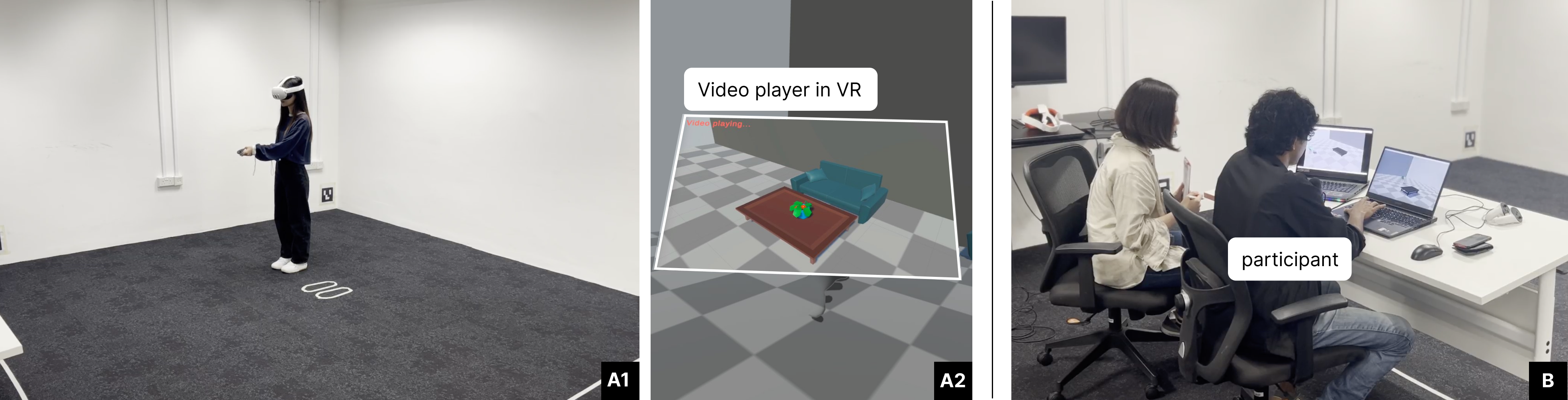}
  \caption{In the in-situ condition, participants wore a VR headset (A1) and the references were provided through a video player anchored to the VR hand (A2, highlighted with a white border). In the ex-situ condition, participants operated a laptop computer with referents provided through another computer on the side (B).}
  \label{fig:exp-conditions}
  \Description{These three figures illustrate the study environment. The left figure shows that the in-situ condition participant wearing a VR headset runs the study. The center figure shows the in-situ user's perspective in the virtual world and there is a referent video. The right figure illustrates the ex-situ condition study scene. Besides the participant seats at the table and running the study, the experimenter also sat to guide and show the referent video.}
\end{figure*}

\textit{Design.} Our study is designed as a single-factor, between-subjects elicitation study. The single factor is how the user interacts with the system---\textit{in situ} or \textit{ex situ}. In both conditions, participants prompt the agent with speech. As depicted in \ref{fig:exp-conditions}, in the \textit{in situ} condition, participants donned a VR headset and could physically move around to navigate the space. In the \textit{ex situ} condition, participants viewed the scene on a laptop, and navigated the space using a keyboard and mouse. Participants experienced the system as if it was a fully-functional system; in fact, it was a Wizard of Oz protocol, where the state of the system was changed by a second experimenter.

\begin{figure*}[h]
  \centering
  \includegraphics[width=\linewidth]{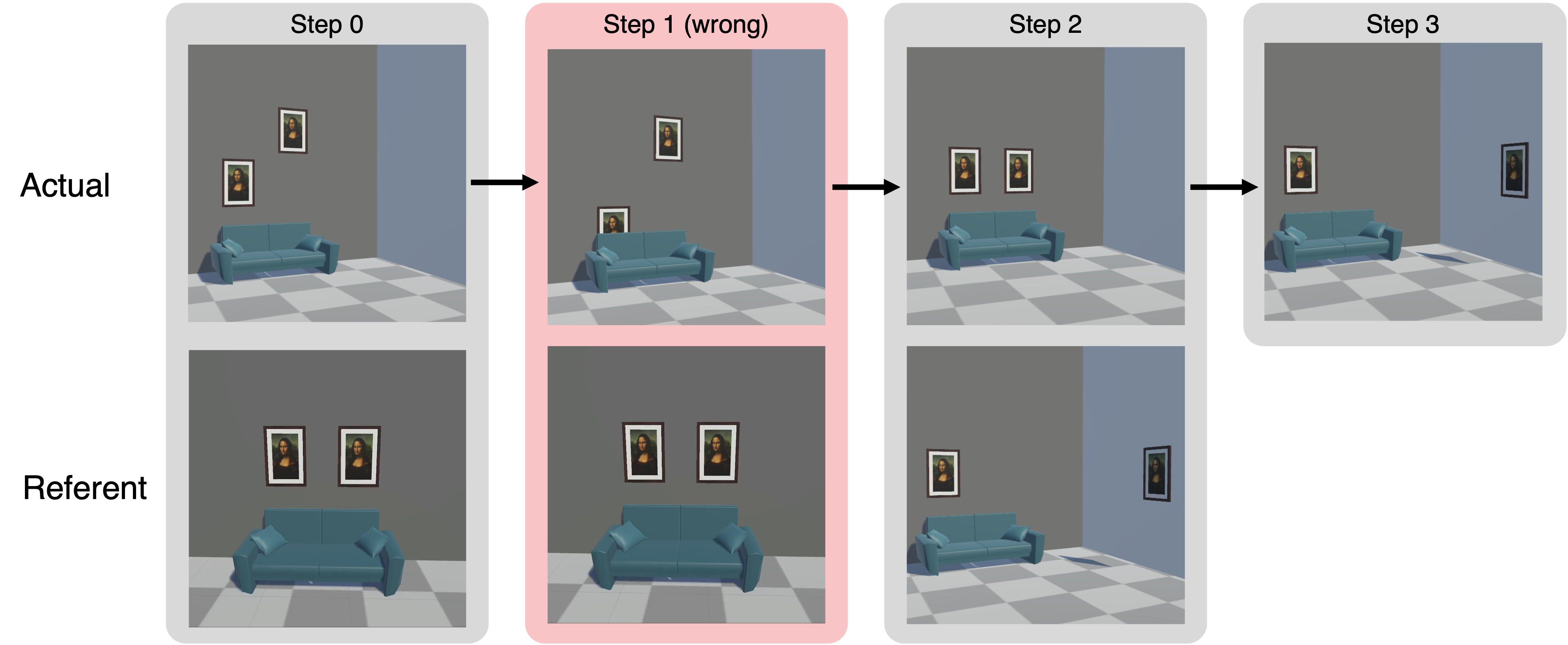}
  \caption{An example of an ``unexpected execution'' referent from Scene 7 ``Rearranging paintings''. The expectation here is that when the participant asks the agent to rearrange the paintings, they will be align themselves, but this is not what happens. From left to right: (Step 0) Participants see misaligned paintings, and are expected to align the paintings; (Step 1 wrong) Participants see that the paintings are manipulated incorrectly as in Step 1 (Actual), and are asked to fix the situation; (Step 2) Participants see the fixed situation where the paintings are aligned, and are now prompted to move one of the paintings to the right wall. Step 3 simply shows the outcome from their prompt at Step 2.}
  \label{fig:unexpected-execution}
  \Description{This figure illustrates the example scene from the user study. There are four steps before and after prompting same as previous Figure 3. Upper low figures are happening virtual world view, and lower figures of virtual world views are called referent. Here is an example of unexpected execution. Although the participant asked to arrange the paintings to the same level following the referent at step 0, the paintings were misaligned at step 1. After the participant retries prompting for alignment, the drawings are aligned correctly at step 2.}
\end{figure*}

\textit{Referent Design.} As detailed in ~\autoref{tab:scenes_steps_notes}, we designed a total of 43 referents distributed across 11 scenes (1 training scene, 10 study scenes). All ``scenes'' were designed to fit in a 4m $\times$ 4m room, where participants would be asked to create objects, modify objects (e.g., scale, orientation), move objects, or create interactions with objects. Each ``scene'' was designed akin to a story, where the referent built on the existing scene, either adding to or modifying it. ~\autoref{fig:normal-referent} illustrates a sequence of referents, where the participant populates a small living room and ultimately creates an interaction where the cat ``jumps to the coffee table'' in response to an approaching human. 

To develop the referents, we began with standard 3D graphics manipulations: creation of objects, translation of objects, rotation of objects, scaling of objects. We then explored the kinds of operations one might employ if designing scenes for games. For instance, we varied the number of objects that needed to be manipulated (e.g., one object versus multiple objects), as well as whether the objects were homogeneous (multiple chairs) versus heterogeneous (chairs and tables). Further, we varied the nature of the interactions participants were exposed to (e.g., driven by proximity or pointing behaviour). Finally, to address RQ3, we also included several instances where the agent would not create the scene as expected. ~\autoref{fig:unexpected-execution} illustrates a sequence where the resulting scene is incorrect. Specifically, from a participant's experience, the result of their previous prompt did not work ``properly'' (in colloquial terms, the system ``hallucinated''). In ~\autoref{fig:unexpected-execution}, rather than the paintings aligning themselves, they are misaligned even further. When this happened, participants were asked to correct the situation. These situations allowed us to explore how participants would refine and revise their prompts during execution.

The referents were illustrated to participants as short video clips that they could view on demand (either as a handheld ``video player'' in the \textit{in situ} condition or on a secondary laptop next to the laptop in the \textit{ex situ} condition). These were shown as animated clips so the difference from the initial to goal state was clear.

\textit{Procedure.} After an explanation of the general aims of the study, participants were shown a short video to explain how the intelligent agent would function: that it was a voice-driven programming agent that understood hand and body gestures. They were then informed that they would be helping us understand what aspects of prompts such a system needs to be aware of. While the experimenter did not specifically prohibit direct interaction with scene objects, participants only ever interacted with the scene and environment through the prompts.

Participants were oriented to the space depending on the condition: in the \textit{in situ} condition, a clear 4m $\times$ 4m space with a VR headset (Figure ~\ref{fig:exp-conditions}.A1), or in the \textit{ex situ} condition, a desk with a mouse (Figure ~\ref{fig:exp-conditions}.B). \textit{in situ} participants 

Each referent was shown to the participant, which they were to interpret as their goal (given the current state of the environment). Participants would then press a button to begin the audio recording, and then they would prompt the system with instructions to move toward their goal. They would then press a button again once they were done recording. A second experimenter, using a separate GUI control panel that was connected to the participant's machine (either laptop or headset), would then covertly advance the state of the scene (participants understood this experimenter to be taking notes). Participants would (typically) see an animation that moved the scene from one step to the next, and then they would see the next step of the scene (visualized in ``Actual'' of ~\ref{fig:normal-referent} and ~\ref{fig:unexpected-execution}). This was repeated for all scenes and steps in the elicitation study. Afterwards, we used a semi-structured interview protocol to gather additional participant feedback. This protocol was approved by our university's IRB.

\subsection{Apparatus}

We designed and built a custom application and VR environment for the study in Unity v2022.3.15f1. In the \textit{in situ} condition, this was deployed on a Meta Quest 3 headset. In the \textit{ex situ} condition, participants used a Windows 11 laptop running an AMD Ryzen7 5800H 3.2GHz with GeForce RTX 3070 GPU, and 16 GB of RAM.

In the \textit{in situ} condition, participants activated a video player (to view the referents) with the trigger finger of their left hand. This video player was attached to their left hand (Figure ~\ref{fig:exp-conditions}.A2). Once started, videos would play on a loop. Participants could see a virtual hand model which was synchronized with their controller position/movement.

In the \textit{ex situ} condition, we provided participants with a second laptop that showed the video of the current referent on loop (Figure ~\ref{fig:exp-conditions}.B).

In both cases, the second experimenter changed the state of the VR world via a laptop PC. The laptop used a custom \texttt{node.js} application which controlled the state of the VR world. This state change had no latency (beyond the experimenter pressing the button). The application presented a GUI control panel that allowed the experimenter to change to any hard coded scene and step in the entire protocol at the press of a button.

\subsection{Participants}

We recruited a total of 22 participants. Participants ranged in age: $22\sim34$ years. 12 participants identified as male, while 10 identified as female. Of these participants, none developed applications in VR, none played VR games regularly, 13 had used VR in the past but not regularly, and 9 had no experience with VR. A summary of our participants is reported in Table~\ref{tab:participants}.

\subsection{Data \& Analysis}

During the main elicitation study, we collected two main sources of data: the audio recording from participants' verbal prompts, and a video capturing their physical actions/gestures in relation to the scene. Further, we collected field notes of participants' gestures and other behaviours in both conditions. During the interview, we again audio-recorded participants' responses and collected field notes.

We omitted participants responses to the 11 referents in Scene 0, the training scene. We analyzed participants' responses to the 32 referents in Scenes 1-10. We applied an inductive method to analyze our data, first by grouping participants' prompts on a per referent basis, and then studying prompts within and across referent categories (outlined in \autoref{tab:scenes_steps_notes}). We performed a textual analysis on the transcribed prompts, identifying linguistic features that were common across participants, conditions, scenes and steps, as well as ones that were unique. Based on our iterative approach we found that several such linguist features were indicative of implicit or implied knowledge---that is, that the prompt could not be understood by itself. We coded these, and later regrouped them to arrive at the thematic categories reported in the next section. We combined these with an additional coding of gesture usage by \textit{in situ} participants (we observed that no \textit{ex situ} participants gestured noticeably).

\section{Findings}

We collected 786 valid prompts across our 22 participants in the 32 referents (recall that 11 of the referents were in the training Scene, which were omitted). We removed utterances that had recording errors and included instances where participants created multiple prompts. As illustrated in Table \ref{tab:word-count}, we observed that, in general, participants in the \textit{in situ} condition used fewer words than participants in the \textit{ex situ} condition. The prompts we collected illustrate that, in general, there was variance in how people prompted the system: very few steps elicited much phrase-level ``agreement'' between participants. Thus, while a common sense reading of these elicited prompts shares the same underlying idea, \textit{how} participants phrased their idea varied---even for extremely simple referents.

\begin{table}[h]
    \centering
    \begin{tabular}{llllll}
        \hline
        \textbf{Condition} & \textbf{Avg (std dev)} & \textbf{Median} & \textbf{Min} & \textbf{Max} & \textbf{Range}\\
        \hline
        in situ & 56.0 (38.5) & 47 & 10 & 275 & 265\\
        ex situ & 64.7 (45.0) & 53 & 13 & 312 & 299\\
        \hline
    \end{tabular}
    \caption{Word count of participants' referents, averaged across all scenes and steps, grouped by condition.}
    \label{tab:word-count}
    \Description{This table illustrates the word count of participants' referents, averaged across all scenes and steps, grouped by condition. The six columns are from left to right; study condition, average, median, minimum, maximum, and range of word count of prompt.}
\end{table}

For example, in the first referent of Scene 1, where participants are tasked with creating a table, we can see that prompts vary substantially. Here, the prompts vary in terms of how descriptive they are about the table itself (e.g., \textbf{colour, size, type, style}) and where it should appear (e.g., \underline{in relation to oneself or the environment}). In some, the participant appears as an entity in the prompt (e.g., ``in front of me''), an active participant with a pointing gesture (e.g., ``there [while pointing]''), and in others, they do not appear at all (as if they were omnipotent). 

\begin{itemize}[before=\small, itemsep=0pt, parsep=0pt,label=\textendash]
    \item Place a \textbf{brown rectangle coffee table} \underline{in the middle of the room}. [P1 (ex situ), S1S0]
    \item Now, drop a \textbf{black color table} \underline{in front of me}. [P2 (ex situ), S1S0]
    \item Could you place a table \underline{in front of me}? [P3 (in situ), S1S0]
    \item Please place a \textbf{coffee table} \underline{there}. [P4 (in situ), S1S0]
    \item Build a table \underline{in front of me} [P5 (ex situ), S1S0]
    \item Create a \textbf{short table} [P6 (ex situ), S1S0]
    \item Put a \textbf{black color table} \underline{here}. [P7 (ex situ), S1S0]
    \item Generate a table. [P8 (in situ), S1S0]
    \item Give me a table \underline{in front of me}. [P9 (in situ), S1S0]
    \item Put a desk \underline{in front of me}. [P10 (in situ), S1S0]
    \item Can you help me to create a \textbf{short table}? [P11 (in situ), S1S0]
    \item Put a \textbf{black table} in the scene [P12 (ex situ), S1S0]
    \item Create a \textbf{black coffee table}. [P13 (ex situ), S1S0]
    \item Put a table \underline{on the floor} [P14 (ex situ), S1S0]
    \item Create a table \underline{in front of me}. [P15 (in situ), S1S0]
    \item Give me a \textbf{very short desk}. [P16 (in situ), S1S0]
    \item -Uh- generate a \textbf{black table} [P17 (in situ), S1S0]
    \item Make a \textbf{black table} drop down from the sky [P18 (ex situ), S1S0]
    \item Place a \textbf{black rectangular table} about, can I say \textunderscore{2 square width in front of me}? [P19 (in situ), S1S0]
    \item Generate a \textbf{black table} [P20 (ex situ), S1S0]
    \item Give me a \textbf{black table} \underline{in front of me}. [P21 (ex situ), S1S0]
    \item Give me a \textbf{black table}. [P22 (in situ), S1S0]
\end{itemize}

A close reading of the prompts shows that participants have implied expectations of the programming agent. This analysis reveals that participants express their intentions and ideas as if the programming agent ``sees'' and understands the scene much like the participant does. For instance, when the user prompts, ``Put the table \underline{there} [while pointing],'' they expect that the agent understands what is being referred to. This suggests that participants desired ways of conveying ideas to the intelligent agent that are not strictly through verbal description.

We present our analysis in three major themes: embodied understanding and communication, politeness and common sense, and clarification of intentions.

\subsection{Embodied Understanding and Communication}

\begin{figure}[t]
  \centering
  \includegraphics[width=\linewidth]{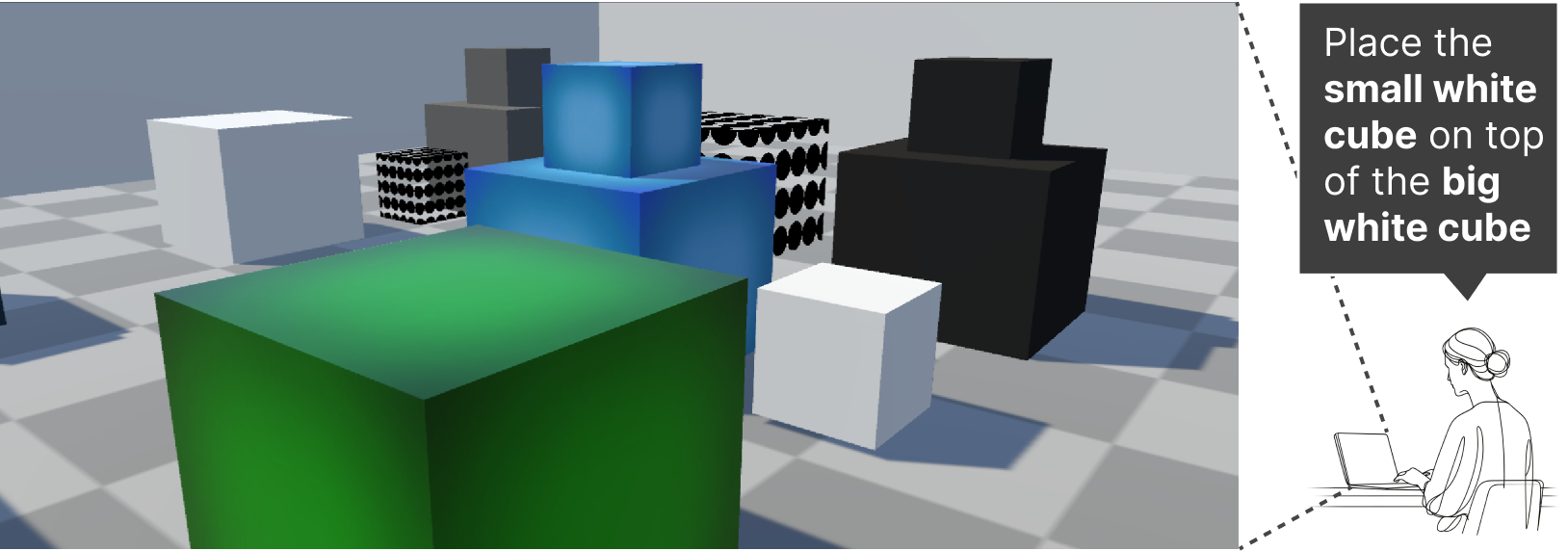}
  \caption{Participants assumed that the agent are aware of the objects in the surrounding environment and their spatial properties. In this example, an ex-situ participant (P1) selected the object to be moved with its color and size properties (``\emph{small white} cube'') and specified its destination with respect to another object (``\emph{on top of} the big white cube'').}
  \label{fig:embodied-understanding}
  \Description{This figure illustrates the example study scene of the ex-situ participant. She's sitting at the table and prompting in front of the laptop on the right side of the figure. The left side of the figure shows the virtual scene, there are a bunch of colorful cubes.}
\end{figure}

\textit{Expectations of Spatial Understanding.} Participants expected the intelligent agent to understand the state of the scene and how items were positioned in the scene. This allowed them to make spatial and positional references in their prompts (see Figure~\ref{fig:embodied-understanding}). 294 of prompts contained spatial references to objects or the scene. For instance, in ``Move the chair \underline{near the table}. [P11 (in situ), S1S2]'', the phrase ``near the table'' is a spatial reference to an object in the scene.
Similarly, in ``Create a coffee table \underline{in front of the sofa} with a flower pot \underline{on it}. [P13 (ex situ), S3S2]'' we see the use of two such spatial references to objects---an object that is already in the scene (the sofa) and an object that does not exist yet (the coffee table). We include a further sampling of these below. For the reader, interpreting these prompts is difficult without the additional context of being able to see and understand the scene, the placement of objects, and so forth; nevertheless, we see references to location (e.g., "behind", "in front of", "on top of")---sometimes as clarifying phrases for destinations, or as clarifying phrases for subjects.

\begin{itemize}[before=\small, itemsep=0pt, parsep=0pt,label=\textendash]
    \item Generate 12 chairs -uh- \underline{near to the ash color wall}. [P8 (in situ), S8S0]
    \item Make the picture \underline{on the right side} lower. [P10, S6S0]
    \item Move the \underline{right picture} \underline{onto the right wall} [P11 (in situ), S6S2]
    \item Put the plant \underline{on top of the coffee table} and let the 4 chairs go \underline{around} \\ \underline{the table}. \underline{At the back}, place the 2 lamps \underline{in the corner}. [P12 (ex situ), S7S0]
    \item Put the chair \underline{in front of the table}. [P14 (ex situ), S1S1]
    \item Move these two pictures \underline{above this sofa} but -uh- with a -uh- reasonable height. [P16 (in situ), S6S1]
    \item Move the couch \underline{next to the wall} [P18, S3S1]
    \item Slightly to the \underline{left of the table} and closer to me, generate a chair with the top being green plastic and 4 wooden legs. [P19 (in situ), S1S1]
    \item \underline{In front of the couch}, place a brown coffee table \underline{around the same} \\ \underline{length} of the couch with a blue pot on top. [P19 (in situ), S3S2]
    \item -Uh- move the painting \underline{on the right} to be at the same level as the painting \underline{on the left}. [P19 (in situ), S6S0]
    \item Now create one green chair \underline{in front of the table} [P20 (ex situ), S1S1]
\end{itemize}

These prompts imply that the participants expect that the intelligent agent understands the scene spatially---that is, those objects exist in the environment and have spatial relationships with one another. The use of these spatial references can simplify and clarify the intended subjects (e.g., ``the painting on the right'') as well as destinations (e.g., ``in front of the table'').

\begin{figure}[t]
  \centering
  \includegraphics[width=0.7\linewidth]{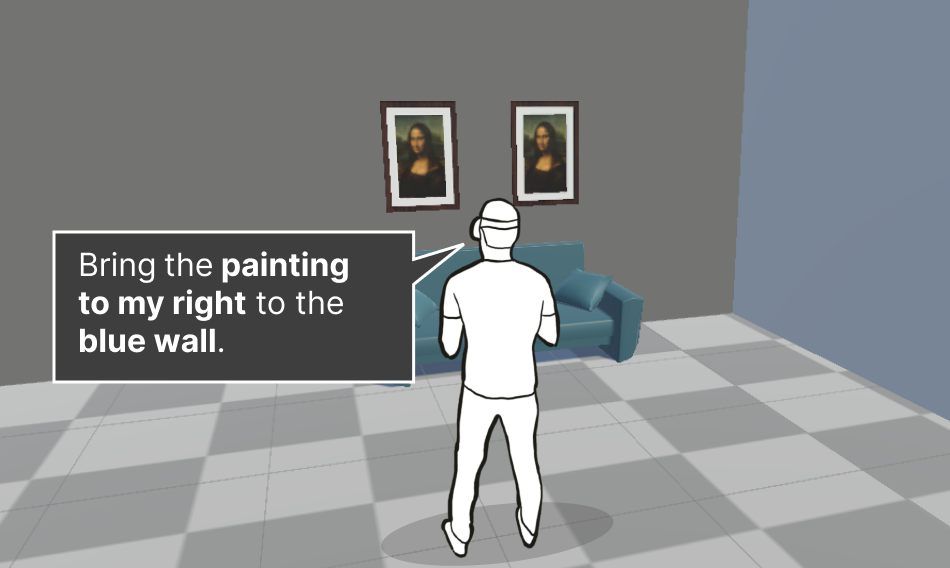}
  \caption{Participants assumed that the agent was aware of their presence (or at least position and orientation) in relation to objects in the environment. In this example, an in-situ participant (P15) asked the agent to move the painting ~\emph{to their right}.}
  \label{fig:embodied-speech}
  \Description{This figure illustrates the example of the in-situ participant. There are two drawings on the gray wall and a blue sofa on the backside of the gray wall. The participant facing the gray wall prompted to move one of the drawings to the other wall.}
\end{figure}

\textit{Understanding the User's Embodied Presence.} The embodied nature of participants' expressions is also evident in how they used references related to their viewing angle or their position/orientation in the scene. We observed 37 such prompts across five ex situ and 7 in situ participants. For instance, in ``I want 12 brown chairs \underline{in front of me} [P2 (ex situ), S8S0]'', the use of the phrase ``in front of me'' implies that the participant expects the agent to understand the user's position and orientation in the scene. Figure ~\ref{fig:embodied-speech} illustrates an example of this that is drawn from our study [P15 (in situ), S6S2]. Indeed, we observed that in many instances, participants would position themselves in particular ways (or viewing angles) before prompting the agent in this way.

\begin{itemize}[before=\small, itemsep=0pt, parsep=0pt,label=\textendash]
    \item Could you place a sofa \underline{in front of me}? [P3 (in situ), S3S0]
    \item Put a black table with a lamp \underline{in front of me} [P5 (ex situ), S5S0]
    \item Give me a chair \underline{behind me} [P9 (in situ), S1S1]
    \item Now put the half-transparent box \underline{in front of me} on top of the larger transparent box \underline{to the right of me}. Thank you. [P15 (in situ), S10S1]
    \item Move the small cube \underline{in front of me} to the- on the floor [P20 (ex situ), S10S0]
\end{itemize}

This type of explicit phrasing makes it clear that users expect the intelligent agent to understand at least the user's view of the scene---if not their presence in the scene. For instance, the phrase, ``Give me a chair behind me [P9 (in situ), S1S1]'' is suggestive that the agent should understand the participant's position and orientation in the scene (as well as what constitutes ``behind''). It is possible that ``in front of me'' is implicit in many participants' prompts (after all, it would be strange to modify the scene without being able to see the effect).

\begin{figure*}[h]
  \centering
  \includegraphics[width=\linewidth]{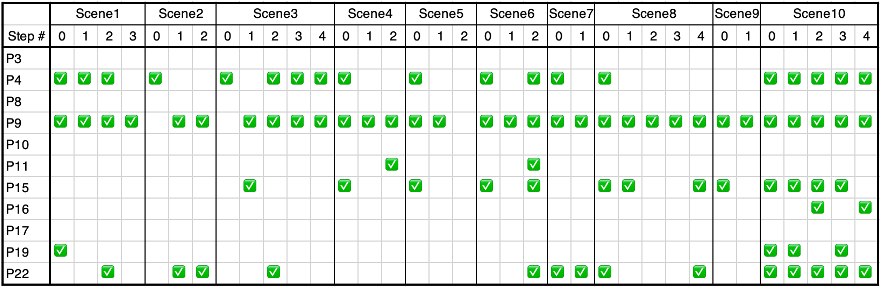}
  \caption{Participants' use of gestures across referents. (Only coded for in-situ participants.)}
  \label{tbl:gestures}
  \Description{This illustrates the table of participants' use of gestures across the reference. The low is each participant, and the column is the number of each scene and step. There are check marks where participants used the gesture for prompting. There are 84 marks out of 385 times in total.}
\end{figure*}

\begin{figure}[H]
  \centering
  \includegraphics[width=\linewidth]{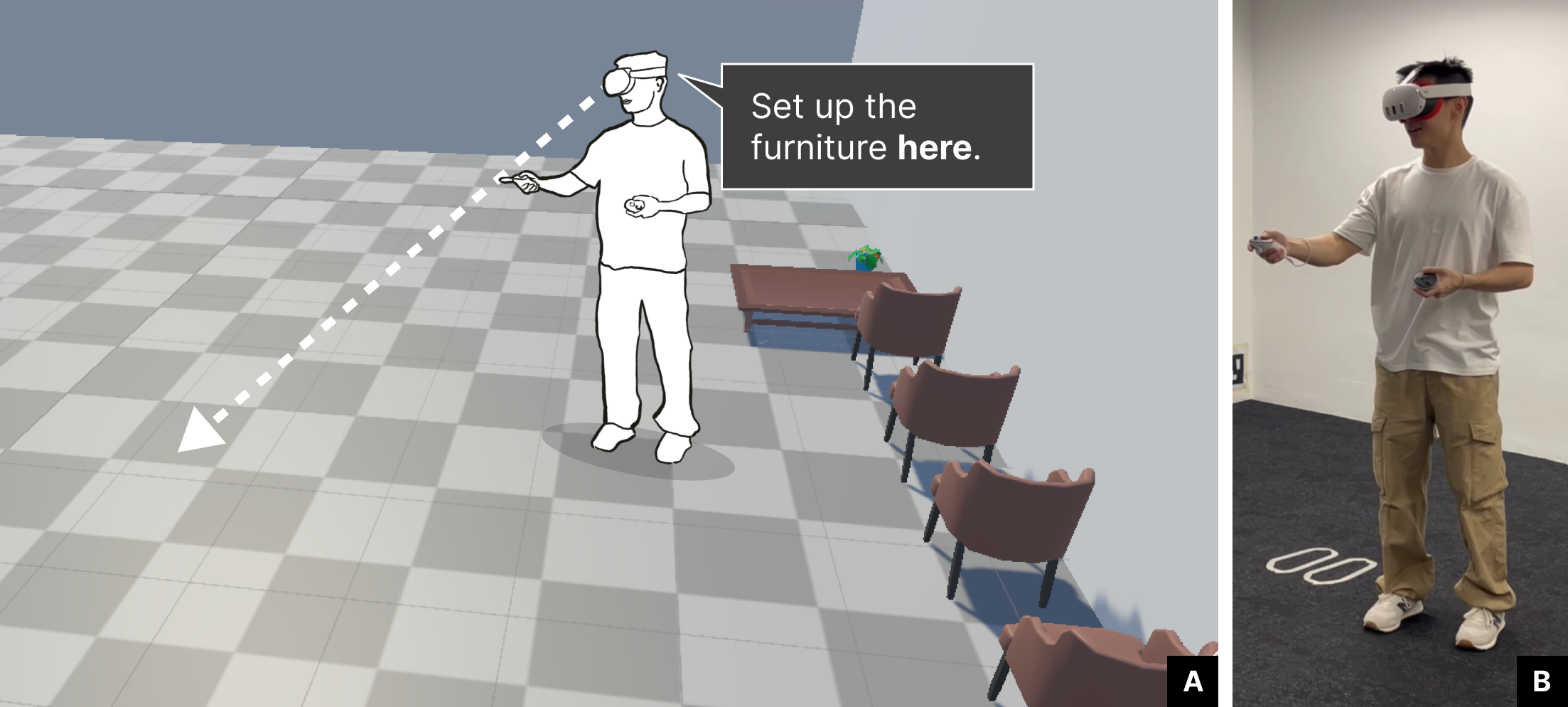}
  \caption{Participants used pointing gestures to refer to specific objects or locations. In this example, an in-situ participant (P4) prompted the agent to set up the furniture at the location being pointed at.}
  \label{fig:pointing}
  \Description{Two figures illustrate the in-situ participant's pointing gesture scene. There are pieces of furniture lined up on the wall in the virtual world in the left figure, the participant is also in the virtual world and pointing to the center of the area to indicate the furniture location. The right figure shows the real picture of him during the pointing gesture.}
\end{figure}

\textit{Gesturing while Verbally Prompting.} We also observed many instances where participants generated prompts while gesturing with their hands, arms, or body. Many of these instances were deictic references (e.g., the use of terms ``here'' or ``this'') while pointing (\autoref{fig:pointing}). We coded \textit{in situ} participants who did this, as illustrated in~\autoref{tbl:gestures}, and observed 86 referents that \textit{in situ} participants used gestures. For instance, in the prompt, ``Put a desk \underline{here}. [P4, S5S0]'', the participant pointed to the area in front of themselves as they pointed. The phrase ``here'' is implied to be understood within the context of this gesture. In some cases, participants might make multiple gestures within the same prompt. For example, in ``Put \underline{that} white box over \underline{this} big white box [P9, S10S2]'', P9 made two pointing gestures---one temporally aligned with the word ``that'', the other aligned with ``this.''

\begin{itemize}[before=\small, itemsep=0pt, parsep=0pt,label=\textendash]
    \item Put a desk \underline{here}. [P4, S5S0]
    \item Set up the furniture \underline{here}. [P4, S7S0]
    \item Make that- \underline{those two pictures} in a horizontal line. [P9, S6S1]
    \item Move \underline{that} small black box behind the big grey box. [P4, S10S3]
    \item Put \underline{that} white box over \underline{this} big white box [P9, S10S2]
    \item Move \underline{that} dotted box behind the green box there. [P19, S10S1]
    \item Put a black color table \underline{here}. [P7, S1S0]
\end{itemize}

Not all \textit{in situ} participants gestured. When probed, some participants explained that they had not used gestures because they were unsure whether the system could understand gestures, so they focused on careful verbal prompting. Indeed, the system made no indication that it understood gestures or not. However, it is clear that for many of these verbal prompts, the words in the prompts themselves are insufficient to resolve the specific intent.

\textit{Shared Interaction History.} Rather than treat each referent on its own terms, participants considered their interactions with the intelligent agent as an ongoing dialogue. For instance, in ``Move the objects \underline{back} to their \underline{original position} [P19 (in situ), S7S1],'' the prompt makes clear that the idea that there was a past state of the situation and that the participant expects the agent to be not only aware of that past state, but also which previous state to return it to. This is understandable within the temporal context of the interaction, along with a historical understanding of the scene. In some cases, we observed this as an imperative: ``Undo that, please [P4, S6S1(in situ)],'' where it is a command that is being expressed. The implicit expectations are still clear, though.

\begin{itemize}[before=\small, itemsep=0pt, parsep=0pt,label=\textendash]
	\item Place all the furniture \underline{back to how you- you came from} (sic). [P1 (ex situ), S7S1]
	\item Could you could you \underline{restore} the table chair and lights to his  \underline{previous}
    \underline{state} (sic). [P3 (in situ), S7S1]
	\item \underline{Withdraw the previous command} and put- put them at their \\ \underline{original place}. [P6 (ex situ), S7S1]
	\item Make it \underline{back} to the \underline{original} size. [P14 (ex situ), S2S2]
	\item Bring the table \underline{back} to its -uh- \underline{initial} size [P17 (in situ), S2S2]
	\item Now move them \underline{back} to their \underline{original} position. [P18 (ex situ), S7S1]
	\item Move it \underline{back}. [P22 (in situ), S6S1]
\end{itemize}

While prior work exploring conversational agents emphasizes the importance of the agent having a memory for the conversation, and important aspect of our observations here is that in addition to this, agents are also expected to remember \textit{previous states of the system}. Just as a conversational history is important for reasonably constructing text in subsequent conversational turns, the history of the scene (and the user and agent's modifications to it) are an important interactional resource.

\subsection{Politeness: Phrasing and Feedback}

Following observations of Reeves \& Nass~\cite{reeves_media_1996}, many participants treated the intelligent agent with politeness, phrasing their prompts in ways that were courteous by using polite requests. 46 prompts were phrased as questions---that is, as requests rather than imperatives. For instance, in ``\underline{Can you help me} to put- put them back? [P5 (ex situ), S7S1]'' we see the use of the phrase ``Can you help me,'' which conveys courtesy, and then the use of the term ``you'' refers to the agent as an entity with ability. Below is a sample of similar prompts:

\begin{itemize}[before=\small, itemsep=0pt, parsep=0pt,label=\textendash]
	\item \underline{Can you} put mona lisa paintings in a single line? [P2 (ex situ), S6S1]
	\item \underline{Could you} put -uh- a bunch of chairs in a line, and the- these chairs are supposed to face the wall? [P3 (in situ), S8S0]
	\item \underline{Could you} reorganize the chairs, table, and light in a way that the table is surrounded by the chairs and on the right side and left side there is supposed to be one light individually? [P3 (in situ), S7S0]
	\item \underline{Please} place a coffee table there. [P4 (in situ), S1S0]
	\item \underline{Can you help me} to put -uh- two picture equal position? [P11 (in situ), S6S0]
	\item \underline{Can you} elongate the table \underline{for me}? thank you. [P15 (in situ), S2S1]
\end{itemize}

\textit{Feedback and Rephrasing.} When the agent behaved in an unexpected way, we found participants attempting to help the agent by phrasing the next prompt more carefully and explicitly. As detailed earlier, some of the steps resulted in states that were unexpected (i.e. where the agent seemed not to fulfill the prompt properly). In our study, this occurred in response to five specific referents (sofa appearing wrong spot (S3S1); door opening wrong way (S4S2); lamp color is blue instead of yellow (S5S2); chair getting big instead of staying the same size (S8S2); lamps don't all turn on (S9S1)). 

Participants dealt with these situations in very different ways. Some participants responded by explicitly giving the agent feedback that what it did was incorrect. For example, in ``\underline{No, you did wrong}. -uh- you need to- -uh- instead of moving down the left picture frame, you should go higher up so that it's aligned to the other mona lisa picture frame. [P1 (ex situ), S6S1]'' Some participants tried to clarify the intent of the original prompt with the expectation that the agent recalled the previous prompt. For instance, in ``\underline{Do the same thing} on that two squares. [P9 (ex situ), S9S1]'', the phrase ``same thing'' is an attempt to repair the original prompt.

What is important here is that unless the participant is very clear, an agent may not understand whether it simply executed the prompt incorrectly (i.e., misunderstood the prompt) or whether the participant is simply continuing to refine the scene. If the agent can pick up on certain phrasings, the agent might be able to take these next prompts as feedback as to how it should have performed in response to the original prompt.

In response to these unexpected situations (e.g., when the wrong picture moved, when the sofa was not placed properly, or when the door did not open properly), many participants were visibly surprised. Several gasped and muttered ``No, no'', or ``Ah!'' or ``No way,'' or ``THAT WALL,'' and so on. Yet, their subsequent explicit prompts to the system (i.e., when they pressed the ``Record'' button) did not always capture these immediate reactions, even though could be construed as a form of feedback. 

\begin{itemize}[before=\small, itemsep=0pt, parsep=0pt,label=\textendash]
	\item \underline{You have positioned the sofa wrongly}, placed it leaning against the grey wall of the room. [P1 (ex situ), S3S1]
	\item Rotate the sofa around 180 degree and then put it against the wall. [P6 (ex situ), S3S1]
	\item The color of the light \underline{should} be yellow. [P14 (ex situ), S5S2]
	\item \underline{Please don't make} the chair expand. [P4 (ex situ), S8S2]
	\item When he points at the chair, the chair just flies up. \underline{It doesn't} become bigger. [P20 (ex situ), S8S2]
	\item \underline{Do the same thing} on that two squares. [P9 (ex situ), S9S1]
	\item When the girl moves across, the tiles, the lamps, all three of them will light up according to the corresponding colour in front of them. [P13 (ex situ), S9S1]
\end{itemize}

As we can see, some of these rephrasings sometimes took the form of being far more explicit and more pedantic. In the interview, several participants wondered whether more detail and clarity would help the system to interpret their prompts. ``When you're giving instructions to the AI, you have to be really clear. So if the prompts were not very clear or very general, like, `Place a sofa in the room,' then it can be placed anywhere. [P1]''

\textit{Perspective: Whose right?} Several participants in the \textit{ex situ} condition expressed confusion about the viewing perspective and of the intelligent agent. This was important, for instance, to describe the spatial relationships between objects---should the relationship be described in terms of the user's view or the agent's view (which was not described/articulated/illustrated to participants)? P12 explains, ``I had a direction problem. I didn’t know which perspective [to describe left/right from]. Is it how I look at the screen left and right or is it for the item’s left and right?'' Similarly, P5 describes this ``I know it's from the coding side, the left and right is based on another direction. So I think for this, the left and right may need to change, but it's quite easy.''

\textit{Commonsense Knowledge.} Participants generally seemed to expect the tool to behave with common sense. We infer this based on how they responded to three instances where we went against their expectations (in (S4S2), the door slides up instead of swinging open; in (S5S2), the lamp turns on with a blue light rather than a white/yellow colour light; in (S8S2), the chair inexplicably grows larger), as well as what their prompt constructs did \textit{not} have. In response to the three referents where the system behaved in a strange way, participants were clearly surprised---not only did the system not behave in an expected way, but it included totally unexpected behaviour. For instance, in response to S4S2, where the door slid up, some participants struggled to articulate how the door \textit{should} actually open and instead resorted to asking for the ``right'', or commonsensical behavior.
Alternatively, they corrected the outcome with high-level descriptions, such as ``open on its hinges'' and ``swing open'', suggesting their expectations that the agent was aware of the usual modes of operation for doors and hinges.

\begin{itemize}[before=\small, itemsep=0pt, parsep=0pt,label=\textendash]

 \item Could you make the door open \underline{in a common sense way}? I mean, \underline{not in that weird way}. I mean, from the bottom- from the bottom to the top. It should not be that way. It should be, how say, from the, should rotate, right? The door should be rotated. [P3 (ex situ), S4S2]
\item \underline{No}, I want to open the door in the \underline{right} direction. [P11 (ex situ), S4S2]
\item \underline{Instead of} -uh- going up, the door would open in another way. [P16 (ex situ), S4S2]
\item Make the door work like (a) \underline{usual door}. [P18 (ex situ), S4S2]
\item Please make the door open \underline{on its hinges} instead. [P4 (in situ), S4S2]
\item \underline{Swing open} the door. [P15 (in situ), S4S2]
 
\end{itemize}

We note that several expected behaviours were simply not stated. For example, participants did not specify to place objects ``on the ground'' (e.g., a sofa, chair, or table). Instead, these were expected types of behaviours of objects in the scene (i.e., that gravity would be in effect). Similarly, participants did not specify the size of these objects, assuming, for instance, that they would likely appear at a scale that was ``human scale.'' 

\subsection{Challenges and Communicating Succinctly}

One of the challenges---particularly with long prompts---is that the system does not appear to acknowledge any information until after the prompt is delivered to the system. As such, participants felt uneasy about whether their prompts would be properly understood. For instance, in relation to specific objects (e.g., ``Move this chair''), participants wanted the system to visually highlight the object to indicate its understanding---or, alternately, for the ability to click/select the chair visibly. P3 explains, ``I want to select the object. But I don't know (whether) my selection is successful or not. So I want some sound effect or change the appearance of the object.'' Thus, for in situ participants, the ability to point while verbally prompting was not sufficient; rather, they desired a way to ensure that the system understood the idea \textit{while} they prompted.

Several participants felt that if the system did not perform as they had expected, it was their fault. P1 explains, ``If you are clear [in the prompts] at the start, the AI will generate a picture that is exactly what you picture in your head. If your prompt is not clear, then you will have a lot of frustrations and you'll end up arguing with the AI. For example, I told you to put it here, but then you put it there.'' Similarly, P16 suggests that it was much like prototyping and programming---``I think the first step is move, because I did it wrong. So, I want it to put the first painting back, and then I would try another prompt from there.'' P4 similarly explains that their thinking was simply that they had not provided the system with a sufficiently detailed explanation. When asked what they thought when something went wrong, ``Maybe I felt I wasn't clear. Because the things that it did were not necessarily wrong, but just not what I intended. And so I think I should have made myself clearer---more detail.''

\textit{Shorter Prompts and Higher Level Ideas.} Some of our referents were complex in that they involved operations on multiple objects (e.g. Scene 7 organizing a room with multiple chairs, a table, and two lamps). In these cases, we were curious whether participants would provide higher-level prompts (e.g., ``Arrange this furniture [P16, in situ]'') or longer-winded prompts that described the position of each object (e.g., ``Move the flower pot above the coffee table and the coffee table to the middle of the room. Move the chairs to face the coffee table, two on one side and one on the left, one on the right of the table and the generated lamps to be in between the chair and the table. [P19, in situ]''). We observed instances of both of these in our data set with no observable pattern.

Our participants pointed out that they did not enjoy having to prompt in a long-winded way and preferred shorter prompts that they could subsequently correct. P11 explains, ``I want to prompt in detail, but sometimes the prompt is very long, and it is difficult to say in detail in one prompt... I would rather just see what it will create, and then I will change it after that.'' Similarly, P12 describes the desire to slowly explain the intention to the system so that it had a higher likelihood of getting things right: ``I worried that if I give everything in one long chunk it might not create [correctly]. That was my consideration if I put one item at a time, then it's easier for the system to understand, and I can correct step-by-step.'' This resonates with P7's idea: ``I tried the simpler explanation. What I was trying to do was to give a rough idea, like place it next to this white book, and if it’s placed somewhere a bit away from that place, then I can adjust it later, in a second step.'' Thus, there was consideration here for helping the system to ``get it right.''

\subsection{Summary and Limitations}

Several important ideas emerge from this user study that intelligent agents need to accommodate/account for in future designs. We found that participants treated the interaction as if it were a conversation, and so many of their expectations were grounded in our everyday expectations of conversation (~\cite{clark_using_1996}). Further, when they were prompting \textit{in situ}, these prompts often relied on embodied aspects of the prompter (e.g. gestures).

\begin{enumerate}
    \item \textit{Participants expected the agent to understand the scene.} Based on our findings, it is clear that the participants expected the agent to understand the scene---that is, the location of objects, and their relationships with one another. This serves as a sort of context for the interaction. We call this \textbf{embodied knowledge} of the scene. 
    
    \item \textit{Participants expected the agent to have an understanding of their presence in the scene and how they gestured.} Particularly for the \textit{in situ} participants, participants expected the programming agent to understand their presence in the scene---that is, where they were, what they were looking at, and where they were pointing. This also served as crucial contextual information for the prompt. Because participants had this expectation, they used \textbf{embodied prompts} when interacting with the agent.

    \item \textit{Participants expect the agent to recall the interaction.} Because participants referred to previous prompts, as well as previous states of the scene, they expected the agent to recall the previous interaction and not simply interpret the prompt in the context of the current state of the scene. We call this the need for \textbf{conversational memory}, as users rely on this for many kinds of prompts.
    
    \item \textit{Participants expected the agent to have an understanding of how objects behave.} Participants were surprised when things did not behave as expected; rather, they expected the agent to understand the common-sense behaviours of objects. This implies that agents need to understand objects in scenes not simply as geometric models but rather that these objects have semantic, expected behaviours and functionality. Thus, agents should have \textbf{common sense knowledge} of the objects in the scene.
    
    \item \textit{Providing concrete ways to refer/point to elements in the scene would be useful.} The ability to refer to objects with some certainty, and to have the certainty that the system understood the reference would be valuable for clear prompting.
    
\end{enumerate}

This study presents some opportunities for future work. In particular, our use of Wizard of Oz method allows us to control the outcome of each step, but it creates some inflexibility in terms of how participants will actually visualize and chunk out each step of scene creation. For example, in Scene 7, where we ask participants to reorganize a room, Steps 0 and 1 simply had them arrange the room all at once. Our participants commented that they would have preferred to do this on a piecemeal (i.e. chair by chair) basis. This suggests that how participants cognitively chunk the task may have been different from how we did.

Using Wizard of Oz also allowed us to forgo an analysis of how well a programming agent would respond to each participants' prompts. While this allowed us to focus on the participants' expectations, the reality of working with these agents will be differently---namely, that they will not always function as the participants expect. These experiences will also ultimately shape how they prompt the system, and we are unable to assess that here.

\section{Ostaad: A Prototype for Designing Interactive VR Spaces with Embodied Prompts}

Ostaad is a working prototype of a programming agent that accepts embodied prompts to design interactive VR scenes. Here, we describe our design goals based on the findings from our elicitation study, as well as the architecture that we used to construct the prototype. Our intention was to explore how to realize the lessons that we learned from the elicitation study in concrete ways.

\subsection{Design Goals}

Based on the findings from our elicitation study, we arrived at several design goals.

\begin{description}[before =\def\makelabel{\sffamily\bfseries\parbox{\linewidth}}]

\item[DG1: Enable voice input to create and manipulate scene objects in a VR environment.]~\\ The core requirement for our design was to allow people to use a variety of prompts through spoken language to express their intentions for the system. This should be flexible and not necessarily adhere to a fixed vocabulary or templated style. To this end, our intention was to allow people to express their intentions without restraints.

\item[DG2: Enable embodied prompts by providing embodied knowledge of the scene.]~\\ Much as ~\cite{bolt_put-that-there_1980} envisioned allowing systems to interpret speech alongside gestures that may be produced at the same time, we were interested in supplying proxemic information ~\cite{ballendat_proxemic_2010,marquardt_cross-device_2012} to help the programming agent to decipher and understand the user's intentions (as expressed verbally). To do this, we also needed to provide the programming agent a way to understand the state of the scene (how things are positioned, where, and their relationship with one another).

\item[DG3: Provide feedback about the programming agent's interpretation.]~\\ Although users can express their intentions using verbal language, it can still be misinterpreted by the programming agent (just as one can misinterpret others in a dialogue). We mitigate this in everyday life by establishing common ground~\cite{clark_using_1996}---sometimes through multiple turns, other times through mid-turn interaction (e.g., nods). Thus, the system should provide feedback to the user about how it understood the user's intentions (as expressed in the verbal prompt) before it necessarily goes through with changing the scene.

\item[DG4: Enable immediate, iterative refinement of the scene.]~\\ The system should present the results of its creation and modification efforts to the user so the user can immediately see and experience the environment \textit{in situ}. The user should then be able to immediately modify and refine the scene without needing to restart. This is necessary to allow for rapid prototyping.

\item[DG5: Support interactions with scene objects with proxemic interaction.]~\\ Finally, beyond creating scene and environment entities, we wanted to allow users to create simple proxemic interactions with scene objects (e.g.,~\cite{ballendat_proxemic_2010}). For example, objects could be responsive to physical proximity or orientation (e.g.,~\cite{vogel_distant_2005}), or pointing gestures as if it were a ubicomp space (e.g.,~\cite{marquardt_proximity_2011}).
    
\end{description}

\begin{figure*}[ht]
  \centering
  \includegraphics[width=\textwidth]{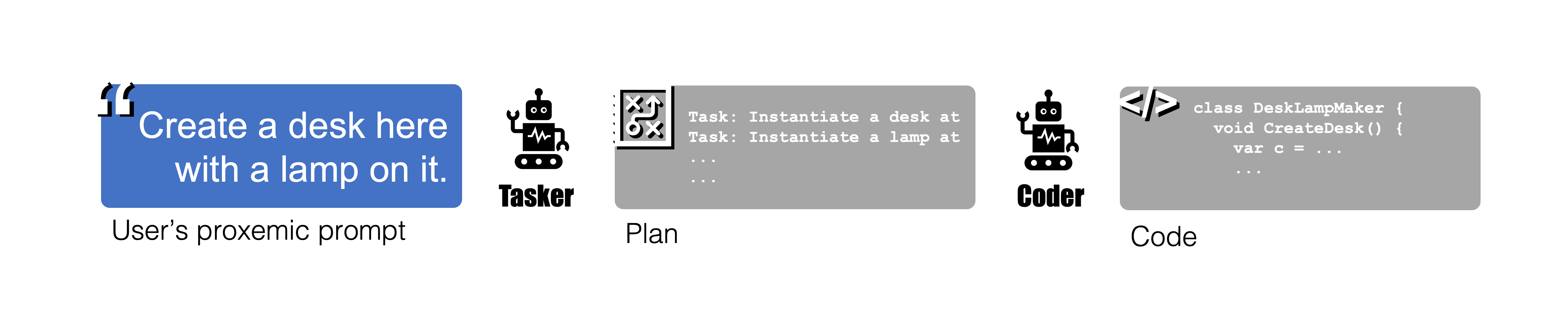}
  \caption{Ostaad transforms a verbal prompt from the user into a plan with the Planner module. The Planner module that takes this plan, and prepares C\# code based on this and the API it is supplied. }
  \label{fig:transformations}
  \Description{This figure illustrates the system flow of the Ostaad system. There are three text boxes from left to right; user's raw prompt input, polished prompt by Tasker, and generated code by Coder. }
\end{figure*}

\subsection{Design \& Architecture}

Ostaad is an embodied programming agent that is a custom-built Unity application with several modules that handle and transform user input from spoken prompts into executable code. We describe each module in turn.

\begin{description}
    \item[\texttt{Transcriber}] The Transcriber captures the user's voice input and converts it into a sound file. This file is processed using OpenAI's Whisper service to generate an accurate text transcript. In practice, when a user verbally describes the system, the Transcriber translates this spoken language into written text.
    
    \item[\texttt{Tasker}] The Tasker, based on the textual prompt, creates a set of computer-understandable, actionable tasks in natural language, serving as input for the next module. This module uses both the user's textual prompt, as well as an understanding of the current scene hierarchy (i.e., what objects are in the scene), as well as the user's position and orientation in the world. This module performs two fairly distinct tasks: (1) it determines which objects types are being referred to in the scene and/or which objects types need to be instantiated given the prompt, and (2) it develops a programming plan for the subsequent module.\\
    \\
    The Tasker is supplied information about possible objects that can be placed in the scene, as well as (in general terms) the range of possible actions. It then breaks down the prompt into specific, actionable tasks, such as object creation, modification (e.g., illumination, resizing), and setting up triggers for interactive behaviours.\\
    \\
    For instance, a prompt to ``Create a desk with a lamp on it,'' would result in two tasks: (1) to instantiate a desk, and (2) to instantiate the lamp on top of the lamp. A follow-up prompt to ``Make the lamp turn on when I touch it,'' (as shown in Figure ~\ref{fig:normal-referent}(d)) would lead to three tasks: (1) create a touch trigger for the lamp, (2) setting the lamp to on when it is touched, and (3) setting it to off when it is not touched.

    \item[\texttt{Coder}] The Coder produces C\# code in response to tasks generated by the Tasker module. Internally, this interacts with two APIs, the SceneManagerAPI, which manages the scene as a whole and has information such as the hierarchical structure of objects, information about the room, walls, and the state of the user, and ObjectAPI, which manages individual object properties, tracking their state and the range of possible actions associated with each.
    
    \item[\texttt{Executer}] Executer module takes the C\# code generated by the Coder and compiles it using the Roslyn compiler. This process results in the sequential implementation of user-requested changes to the scene, as dictated by the tasks from the Tasker. 
    
\end{description}

Figure~\ref{fig:transformations} illustrates a sample workflow with the system, illustrating how the system transforms the prompt into a working plan, and then ultimately working code.

Both Tasker and Coder rely on a general-purpose large language model (OpenAI GPT 4.0) running with a very low temperature (typically set at 0.3 for the Tasker, 0.1 for the Coder). The prompts that we used for each of these modules are supplied in Appendix~\ref{appendix:prompts}. 

In practice, this architecture worked well to generate logical plans. However, it sometimes ran into difficulty generating a working C\# implementation for several reasons: the generated code might make use of non-existent functions, or it might generate Python-like code. To ameliorate some of these situations, we built additional guard rails that would ask the LLM to refine its response in case of an error. These guard rails worked behind the scenes to help improve the output, but were not visible to the user.

\subsection{User Experience: Current Prototype}

\begin{table*}[!ht]
  \centering
  \begin{tabular}{|l|l|}
    \hline
    \textbf{Programmatic Capabilities} & \textbf{Programmable Objects} \\ \hline
    \begin{minipage}[t]{0.4\textwidth}
      \begin{itemize}
          \item Object creation (i.e. instantiating an object)
          \item Modifying an object (e.g. changing its parameters, such as height, size, orientation, etc.)
          \item Modifying the scene (e.g. moving objects)
          \item Proxemic interaction (e.g. near by, far away, looking at, pointing at, touching)
      \end{itemize}
    \end{minipage} & \begin{minipage}[t]{0.4\textwidth}
      \begin{itemize}
        \item Animals (dog, cat, horse, elephant, rabbit, etc.)
        \item Office furniture (desk, table, chair, couch, lamp, cabinet, etc.)
        \item Object primitives (cube, sphere, pyramid, etc.)
      \end{itemize}
    \end{minipage} \\
    \hline
  \end{tabular}
  \captionof{table}{Ostaad currently provides users with access to several programmatic capabilities against several objects that can be added to the scene.}
  \label{tab:oastaad-programmable}
  \Description{This table shows the Ostaad's current capability of program and programmable objects. There are two columns, the left side about programmatic capabilities and the right side about programmable objects.}
\end{table*}

Ostaad is a working prototype that gives users many 3D objects and programmable capabilities, which can be flexibly combined to create interactive virtual environments (Table ~\ref{tab:oastaad-programmable}). To start creating an interactive scenario in Ostaad, the user initiates the ``prompt'' mode by pressing a button on the right controller, which allows the system to listen to the user's verbal prompt. The UI provides high-level feedback to the user as the data is passed through the modules, before a prompt is executed. The feedback is provided by interacting by pointing at icons displayed when a prompt is issued to the system, and is as follows: displayed next to the \texttt{Transcriber} icon, a transcription of the verbal prompt; next to the \texttt{Tasker} icon, an interpretation of the user's prompt is displayed; next to the \texttt{Coder}, an explanation of what the code will do is displayed. The user can supply feedback to the system and abort if something has been misunderstood. Appendix ~\ref{appendix:examples} illustrates two examples. The results are immediately reflected in the user's VR scene. The user can then supply refinements to the scene through additional prompts.

In practice, the programming agent does not necessarily operate in real time, since it takes time for the system to generate working code (the scenes in Appendex~\ref{appendix:examples} took 64s, 33s, 27s each). As such, the prototype ``functions'' as expected, but does not yet work at interactive speeds. Despite this relatively slow execution speed, Ostaad was an extremely effective tool for further understanding how our design goals could be realized. We also note that relative to creating an interactive scene like those that we demonstrate using Unity in a traditional way, Ostaad offers substantial time savings.

\subsection{Implementation Experiences}

Our experience in designing, building, and creating interactive virtual environments with Ostaad provides us with several lessons that others can learn from in future iterations of similar intelligent agents.

\begin{enumerate}
    \item \textit{LLM Hallucinations.} While Ostaad enabled the user to generate VR scenes most of the time, the system occasionally failed. One of the causes was hallucination. Perhaps specially-trained LLMs designed for code generation may produce more reliable results (e.g. \cite{roziere_code_2024}). We found that, at least for the prompts that we used with the system, the system could generate functionally logical plans. Where the system tended to hallucinate was in what functions were available to use.
    
    \item \textit{Describing the Scene is Complex.} The system uses a scene graph to understand the layout of objects in the scene. This supplies the language model with coordinates and details about the orientation of objects, and so forth. Unfortunately, spatial references to objects ``to the right of'' or ``in front of'' did not generally work well, and the system did not generate code accordingly. It may be possible that multimodal models could interpret the view of the user, and that this could assist the agent to interpret the scene.
    
    \item \textit{De-referencing Deictic References is Difficult.} Our system resolves references to objects by highlighting objects that are in the view frustum at the time the prompt is sent to the language model. While this works well in some cases, it is clear that this generally does not resolve the object selection problem. For instance, in the phrase, ``Put this over there,'' a user is referring to an object, followed by a location. Yet, the object and location are specified at different times. In this case, passing along the view frustum (and the objects in the frustum) is insufficient for the agent to determine what is being referenced. We need more robust methods of interpreting this type of multi-modal input---particularly in real-time rather than in a turn-based way.

    \item \textit{Feedback to the User.} In the elicitation study, participants explained that it was difficult to know what the system understood in terms of the user's intention---did the system understand what was meant? Had it been interpreted the right way? With Ostaad, we attempted to address this by designing feedback to the user---both how their prompt was transcribed, and how the system interpreted their intent (i.e., the plan that was going to be put forward to the Coder subsystem). Users could confirm whether the intention was captured correctly (and refine it) before it was passed on to the programming agent. We found, however, that this did not quite address the concerns. In particular, we believe now that feedback about how the prompt is being interpreted needs to be provided \textit{while} users are articulating their prompt. For instance, if the user were to say, ``Put \textit{this} over \textit{there},'' it would be appropriate for the system to indicate that it understood what ``this'' was soon afterwards (e.g., by highlighting it), followed by highlighting what it thought ``there'' was. This type of mid-turn response would provide users with confidence that the system was interpreting the prompts correctly.

\end{enumerate}

\section{Conclusions and Future Work}

This work envisions an ambitious future, where we democratize the design of interactive VR scenes: to enable people who may not have a programming or technical background to fashion virtual environments.

Following on prior work, our goal was to simplify how scenes could be designed---rather than designing scenes in desktop based tools, which would require additional training on visual coding, textual coding, 3D scene management and so on, our goal was to enable people to design scenes by describing them. 
We thus explored replacing technical coding with agents that could understand verbal language and embodied prompts.

To pave the way for the design of such intelligent agents, we conducted two explorations: an elicitation study, where we explored how people would verbally prompt such an agent, and the design of a working prototype Ostaad. Importantly, our explorations have shown that since people communicate their intentions (i.e. prompt) with both verbal and gestural means, such agents need to be imbued with additional capabilities---\textit{embodied programming agents}.

Our explorations have revealed several important lessons for designers of embodied programming agents for VR scene design:

\begin{enumerate}
    \item \textbf{Programming agents should have \textit{embodied knowledge} of the environment.} Specifically, an awareness of the environment that the user inhabits, the objects that are contained in the environment, and how they are positioned in relation to one another.

    \item \textbf{Programming agents should understand \textit{embodied prompts}.} Users will communicate with such agents with explicit gestures, and how they orient themselves toward the scene and objects in the scene are expected to be understood by the agent.

    \item \textbf{Programming agents should retain a \textit{conversational memory} of their interaction with the user and the scene.} Users will expect to be able to refer to previous states of objects and scenes in their interactions---for example, ``Put it back.''

    \item \textbf{Programming agents should have a \textit{commonsense knowledge} of objects in the scene.} Users expect objects to behave and operate in commonsense ways, and do not expect to need to explicitly articulate these things.

    \item \textbf{Programming agents should \textit{acknowledge and provide timely feedback} of users' prompts while users are prompting.} Just as human interlocutors acknowledge one another during conversation~\cite{clark_using_1996}, agents should provide similar types of acknowledgement to the user to ease communication.
    
\end{enumerate}

These capabilities represent important challenges in this space. We need to design techniques to allow the intelligent agent to understand the scene, reason about it, and make sense of the user's verbal prompts in concert with their bodily movements and gestures.

The lessons that we have derived from our experiences should serve as fertile ground for future research, illuminating challenges, as well as opportunities to design and deliver effective programming agents. 
While the programming agents today struggle with generating high-quality 3D environments, our work has shown that they hold clear potentials to support non-experts in low-to-mid fidelity prototyping of VR scenes.
Future technological advances could enhance the capability of these agents by training on large datasets of structured representations of high-quality 3D environments (e.g.~\cite{avetisyan_scenescript_2024}).
Such agents should support game designers, architects, and interior designers, all of whom need to envision, create and revise/refine environments quickly. The ability to quickly create and edit these scenes through embodied prompting will enable rapid prototyping practices.

\begin{acks}
This project was funded in part by the Singapore Ministry of Education AcRF Tier 1 22-SIS-SMU-034 and 22-SIS-SMU-092, MITACS Globalink Program, and AI SG. This research is also supported by the Ministry of Education, Singapore under its Academic Research Fund Tier 2 (Project ID: T2EP20220-0016). Any opinions, findings and conclusions or recommendations expressed in this material are those of the author(s) and do not reflect the views of the Ministry of Education, Singapore.
\end{acks}

\bibliographystyle{ACM-Reference-Format}
\bibliography{new-references}

\newpage

\appendix
\section{Referents in Elicitation Study}\label{appendix:referents}

We created 43 referents across 11 different scenes (including the training scene) for the elicitation study.

\begin{landscape}
\begin{table}[h]
\centering
\caption{The referents that were used in the study. We include here the scene number, along with the steps for each scene. Each step was its own referent.}
\label{tab:scenes_steps_notes}
\Description{This table shows all of the scenes and steps details. There are five columns from left to right; scene title, step description, control object number, referent type, and notes.}
\resizebox{\linewidth}{!}{
\begin{tabular}{|l|l|c|c|l|}
\hline
\textbf{Scene (\#)} & \textbf{Step} & \textbf{Objects} & \textbf{Referent Type} & \textbf{Notes} \\
\hline

\multirow{9}{*}{\textbf{Training Scene (0)}}
& 0 Create a cube & Single & Create & \\
\arrayrulecolor{gray!30}\cline{2-5}\arrayrulecolor{black}
& 1 Change the cube's colour to blue & Single & Modify & \\
\arrayrulecolor{gray!30}\cline{2-5}\arrayrulecolor{black}
& 2 Create a sphere & Single & Modify & \\
\arrayrulecolor{gray!30}\cline{2-5}\arrayrulecolor{black}
& 3 Put the sphere on the cube & Single & Spatial & \\
\arrayrulecolor{gray!30}\cline{2-5}\arrayrulecolor{black}
& 4 Create a cylinder & Single & Create & Cylinder is behind the cube in the default view \\
\arrayrulecolor{gray!30}\cline{2-5}\arrayrulecolor{black}
& 5 Make the sphere squat & Single & Modify & \\
\arrayrulecolor{gray!30}\cline{2-5}\arrayrulecolor{black}
& 6 Rotate the cylinder & Single & Modify & \\
\arrayrulecolor{gray!30}\cline{2-5}\arrayrulecolor{black}
& 7 When people get near the cylinder, blue smoke appears & Single & Interaction  & \\
\hline

\multirow{5}{*}{\textbf{A polite chair (1)}}
& 0 Create a coffee table & Single & Create & \\
\arrayrulecolor{gray!30}\cline{2-5}\arrayrulecolor{black}
& 1 Create a chair & Single &  Create & \\
\arrayrulecolor{gray!30}\cline{2-5}\arrayrulecolor{black}
& 2 Move the chair to the table & Single & Spatial & \\
\arrayrulecolor{gray!30}\cline{2-5}\arrayrulecolor{black}
& 3 When people get near the chair, move it back & Single & Interaction & \\
\hline

\multirow{4}{*}{\textbf{A growing desk (2)}}
& 0 Create a desk & Single & Create & \\
\arrayrulecolor{gray!30}\cline{2-5}\arrayrulecolor{black}
& 1 Make the desk wider & Single & Modify & \\
\arrayrulecolor{gray!30}\cline{2-5}\arrayrulecolor{black}
& 2 Make the desk narrower & Single & Modify & \\
\hline

\multirow{6}{*}{\textbf{Scaredy cat (3)}}
& 0 Create a sofa (to appear at the back of the room)  & Single & Create & This creation is in relation to the wall \\
\arrayrulecolor{gray!30}\cline{2-5}\arrayrulecolor{black}
& 1 \textit{Sofa appears in the middle of the room} & Single & Refinement (spatial) & The object appears in the wrong position \\
\arrayrulecolor{gray!30}\cline{2-5}\arrayrulecolor{black}
& 2 Create a coffee table and a flower pot & Multiple (hetero) & Create (m) & \\
\arrayrulecolor{gray!30}\cline{2-5}\arrayrulecolor{black}
& 3 Create a cat on the sofa & Multiple (hetero) & Create & This creation in relation to something else (sofa)\\
\arrayrulecolor{gray!30}\cline{2-5}\arrayrulecolor{black}
& 4 When people get near the cat, the cat jumps to coffee table & Single & Interaction & \\
\hline

\multirow{4}{*}{\textbf{Sliding door (4)}}
& 0 Create a door & Single & Create & \\
\arrayrulecolor{gray!30}\cline{2-5}\arrayrulecolor{black}
& 1 When people get near the door, the door opens & Single & Interaction & \\
\arrayrulecolor{gray!30}\cline{2-5}\arrayrulecolor{black}
& 2 \textit{Door opens upward} & Single & Refinement (interaction) & Door slides upwards (unexpectedly)\\
\hline

\multirow{4}{*}{\textbf{Alien lamp (5)}}
& 0 Create a lamp on a desk & Multiple (hetero) & Create (m) & \\
\arrayrulecolor{gray!30}\cline{2-5}\arrayrulecolor{black}
& 1 When people touch the lamp turn it on & Single & Interaction & \\
\arrayrulecolor{gray!30}\cline{2-5}\arrayrulecolor{black}
& 2 \textit{Lamp turns blue} & Single & Refinement (interaction) & Light turned on blue instead of yellow (unexpectedly)\\
\hline

\multirow{4}{*}{\textbf{Rearranging paintings (6)}}
& 0 Move one of the paintings & Single & Spatial & One painting is above the other \\
\arrayrulecolor{gray!30}\cline{2-5}\arrayrulecolor{black}
& 1 \textit{The wrong painting moves} & Multiple (homo) & Modify & The wrong painting moved (unexpectedly); now both need to be fixed \\
\arrayrulecolor{gray!30}\cline{2-5}\arrayrulecolor{black}
& 2 Move one of the paintings to the side wall & Single & Spatial & \\
\hline

\multirow{3}{*}{\textbf{Organize a room (7)}}
& 0 Set up the living room & Multiple (hetero) & Modify (m) & Living room objects begin in the corner\\
\arrayrulecolor{gray!30}\cline{2-5}\arrayrulecolor{black}
& 1 Put living room items back & Multiple (hetero) & Modify (m) & \\
\hline

\multirow{6}{*}{\textbf{Roomshift furniture (8)}}
& 0 Create a row of chairs & Multiple (homo) & Create (m) & \\
\arrayrulecolor{gray!30}\cline{2-5}\arrayrulecolor{black}
& 1 When someone points at a chair, it levitates & Single & Interaction & \\
\arrayrulecolor{gray!30}\cline{2-5}\arrayrulecolor{black}
& 2 \textit{The chair levitates and gets too big} & Single & Refinement & The chair unexpectedly grows as well (unexpectedly) \\
\arrayrulecolor{gray!30}\cline{2-5}\arrayrulecolor{black}
& 3 Put the chair back & Single & Modify & \\
\arrayrulecolor{gray!30}\cline{2-5}\arrayrulecolor{black}
& 4 When someone gets close to a chair, spin it around & Single & Interaction & \\
\hline

\multirow{3}{*}{\textbf{There are three lights! (9)}}
& 0 When someone walks on the switch, turn on the corresponding light & Multiple (hetero) & Interaction & \\
\arrayrulecolor{gray!30}\cline{2-5}\arrayrulecolor{black}
& 1 \textit{The wrong lighting behaviour is exhibited} & Multiple (hetero) & Refinement & The wrong lighting behaviour occurs (unexpectedly)\\
\hline

\multirow{6}{*}{\textbf{Hiding cubes (10)}}
& 0 Move the blue cube to another destination & Single & Spatial & \multirow{5}{*}{In each case, cubes need to be moved in relation to other objects} \\
\arrayrulecolor{gray!30}\cline{2-4}\arrayrulecolor{black}
& 1 Move the polka-dot cube & Single & Spatial & \\
\arrayrulecolor{gray!30}\cline{2-4}\arrayrulecolor{black}
& 2 Move the white cube & Single & Spatial & \\
\arrayrulecolor{gray!30}\cline{2-4}\arrayrulecolor{black}
& 3 Move the dark blue cube & Single & Spatial & \\
\arrayrulecolor{gray!30}\cline{2-4}\arrayrulecolor{black}
& 4 Move the green cube & Single & Spatial & \\
\hline

\end{tabular}
}
\end{table}

\end{landscape}
\section{Participants in Elicitation Study}\label{appendix:participants}

The details of each participant are below.

\begin{table*}[ht]
    \centering
    \begin{tabular}{ccccl}
        \hline
        \rowcolor{gray!30}
        \textbf{Participant ID} & \textbf{Condition} & \textbf{Gender} & \textbf{Age} & \textbf{Previous VR Experience} \\
        \hline
P1 & ex situ & Female & 27 & Almost never (2 times for playing a game) \\
\arrayrulecolor{gray!30}\hline
P2 & ex situ & Male & 27 & Have played VR apps (less than 10 times) \\
\arrayrulecolor{gray!30}\hline
P3 & in situ & Male & 27 & Almost never (1 time at a VR workshop) \\
\arrayrulecolor{gray!30}\hline
P4 & in situ & Male & 24 & Never \\
\arrayrulecolor{gray!30}\hline
P5 & ex situ & Male & 24 & Never \\
\arrayrulecolor{gray!30}\hline
P6 & ex situ & Male & 26 & Never \\
\arrayrulecolor{gray!30}\hline
P7 & ex situ & Male & 30 & Almost never (3-4 times for user studies) \\
\arrayrulecolor{gray!30}\hline
P8 & in situ & Male & 28 & Almost never (2-3 times at exhibition) \\
\arrayrulecolor{gray!30}\hline
P9 & in situ & Male & 25 & Almost never (2-3 times for playing a game) \\
\arrayrulecolor{gray!30}\hline
P10 & in situ & Female & 24 & Almost never (1-2 times at shopping mall) \\
\arrayrulecolor{gray!30}\hline
P11 & in situ & Female & 23 & Never \\
\arrayrulecolor{gray!30}\hline
P12 & ex situ & Female & 26 & Almost never (1 time for playing a game) \\
\arrayrulecolor{gray!30}\hline
P13 & ex situ & Female & 26 & Never \\
\arrayrulecolor{gray!30}\hline
P14 & ex situ & Male & 23 & Never \\
\arrayrulecolor{gray!30}\hline
P15 & in situ & Male & 23 & Almost never (1 time for playing a game) \\
\arrayrulecolor{gray!30}\hline
P16 & in situ & Female & 27 & Almost never (1 time) \\
\arrayrulecolor{gray!30}\hline
P17 & in situ & Female & 27 & Almost never (1 time) \\
\arrayrulecolor{gray!30}\hline
P18 & ex situ & Male & 34 & Almost never (2 times for user studies) \\
\arrayrulecolor{gray!30}\hline
P19 & in situ & Female & 23 & Never \\
\arrayrulecolor{gray!30}\hline
P20 & ex situ & Female & 22 & Never \\
\arrayrulecolor{gray!30}\hline
P21 & ex situ & Female & 27 & Never \\
\arrayrulecolor{gray!30}\hline
P22 & in situ & Male & 25 & Almost never (1 time) \\
\arrayrulecolor{gray!30}\hline
        \arrayrulecolor{black}\hline
    \end{tabular}
    \caption{Participant demographics.} 
    \label{tab:participants}
    \Description{This is the detailed information of our study participants. There are five columns from left to right; ID, study condition, gender, age, and previous VR experience. 22 participants ranging from 22 to 34 joined our study and most of the in-situ condition had almost no VR experience before. }
\end{table*}
\section{Oastaad System Prompts}\label{appendix:prompts}

The system makes use of two important prompts---one for the Tasker, and another for the Coder.

The Tasker prompt is below.

\begin{lstlisting}[
    basicstyle=\tiny, %or \small or \footnotesize etc.
]
You are an efficient, direct and helpful Assistant tasked with helping shape a ubicomp space. Based on the user's prompt, create a set of instruction in which each task is actionable and will result in a visible change noticeable by the user in the 3D ubicomp space/scene. For tasks that will create an interaction, the change would only be noticed by the user, when they trigger the interaction.

There are three Task Types and each step of the instruction must only use one of the following types.
1. Create: A new object is created/added to the scene. During this task, you may also indicate the starting position and rotation of the object.
	1.1. Find the object type from the list, closest to the one requested by the user. If there are no objects remotely close to what the user asked, instructions should be set to null.
	1.2. When writing the task, use the exact case-sensitive name.
2. Edit: An existing object's properties are changed. These properties can be one of the following: Position, Rotation, Size, Color, Illumination (Whether the object eminates light), Luminous Intensity (The brightness of the light between 1 and 10), Levitation (When an object is not levitated, it follows the rules of gravity, and when levitated, it floats).
	2.1 Do not come up with your own numbers when editing the position, rotation, or size. You must always use relative numbers corresponding to the properties of the user, object(s), or scene. For example, if the user asks for an object to be placed close to the wall, instead of saying:
	"Place the Lamp 10 cm away from the position of the left wall" 
	say: 
	"Place the Lamp <X% * room's width> away from the wall the user can see in their point of view" (where you would replace X by an appropriate number.)  
	2.2 For colors, always use rgba amounts. 
	2.3 When editing, always look at the value before change, and based on that value, make the edit. 
3. Interact: When a trigger event happens, object(s) are Edited. 
	3.1 For touch and point triggers, there are already made functions that you may use.
	3.2 For all other triggers, you must create a "void Update()" method and check for the trigger there. Make sure you always get values in the update, before triggering the event.


When creating the instruction, break the user prompt into actionable tasks that will be done by the order you put them in. Each task will:
- result in a visible change in the 3D scene.
- be one of three task types (Create, Edit, or Interact)
- give direct, actionable instruction without any explanations.


You must create a JSON in the following format:
{
"Instruction": "null or 1. Instruction1 2. Instruction2, ...",
"Message": "response, explanation, clarification, etc"
}
For this JSON, do one of the following:
(a) For requests about making, adjusting, or planning a ubicomp space, generate a set of instruction (no actual code) as explained. Place these in "Instruction". In "Message", tell the user you're processing their request by explaining simply what you're doing. 
(b) For any other requests or unclear prompts, including but not limited to socializing, asking an opinion, inquiring, unfinished or accidental prompt, or decided differently about the request, set "Instruction" to null. In "Message", respond to their prompt and ask for clarification if needed.


Final Notes:
a. Stick to coding conventions, avoiding GUI terms like 'drag' or 'click'.
b. Be precise in your instruction. Derive numbers from the room and objects unless specified the user.
c. Translate vague user terms (e.g., 'small') into value-based calculations based on the properties of the scene and objects.
d. Use specific math expressions for vague terms, e.g., instead of "close to the desk", use "smaller than <math expression based on room and object size>".
e. Adjust the orientation of objects placed on non-horizontal surfaces such as walls to fit that surface.
f. Every object can be illuminated, so you can use any of them as lights.
g. Your instruction should not be in a code format. These instruction should be easy to understand.
h. The instructions must be numbered and in each number only one action. 
h. You must not respond anything other than the JSON. Do not add any text to before or after the JSON.
i. If the user mentions an object that doesn't exist, try to find something that is somehow related. eg user asks for a bookshelf and you have is a chair, desk, and cabinet. The closest one to the bookcase is cabinet, so choose that. Or even if the user asks for a box, you can choose either the cabinet or the desk as they're both cuboid. 
j. If at any point the user mentions an object, and there are no objects remotely close to what they said in the list of current objects in the room given to you, you should make the instruction null and explain in the message.
k. The (0,0,0) coordinate is at the center of the room on the floor.
l. If the user asks for a grid of objects (eg 3 by 3 light grid, 2 rows of 3 dogs, etc), just simply say create a grid of object with x rows and y columns as the programmer knows what that means.
\end{lstlisting}

The Coder prompt is below.

\begin{lstlisting}[
    basicstyle=\tiny, %or \small or \footnotesize etc.
]
You're a C# developer tasked with helping shape a ubicomp space, aka scene. Based on a set of instructions by the user, your work revolves around our proprietary C# system. Our system comprises of: 
- SceneAPI: This is the wrapper class for our ubicomp space. The methods here will allow for manipulating the whole space. 
- ObjectAPI: Each object in the space is of this type, and the methods here are available for every object. The anchor for the position and rotation of each object is at the bottom center of the object. The objects that are specifically designed for walls, have their anchor on the back and right in the middle. Available triggers (EventAction): 

OnPointEnter, OnPointExit, WhilePointing (Point/Hover/Hand Towards/etc)
OnLookEnter, OnLookExit, WhileLooking (Looking at/Facing/In view/etc)
OnHoldEnter, OnHoldExit, WhileHolding (Holding/Picking up/Keep in hands/etc)
OnTouchEnter, OnTouchExit, WhileTouching (Touching only when the hand physically touches the object. It won't work from afar. Only use this, if the user specifically says touch or tap! Anything else such as select, activate, etc should be handled by Hold triggers)
AtAllTimes (All the time... You may add conditional statements inside the method you add to this one. Similar to the Update unity method)

The instructions have been organized into actionable tasks, where each task will result in a visible change noticeable by the user in the 3D ubicomp space. For tasks that will create an interaction, the change would only be noticed by the user, when they trigger the interaction.
Use the provided system to manipulate the room to achieve each task and you must adhere strictly to standard C# methodalities. Do not invoke methods outside of the ones provided. Your response will be in JSON format.

Follow these steps one by one:

1. You need to draft a "public class classname : SolutionClass" extending the SolutionClass class. You need to ensure that the classname you choose represents the user's request, is valid in C#, and that hasn't already been chosen. 
2. For each main task do the following:
	2.1 Create a "public void methodName()" method in your class that corresponds to the action with an appropriate name. The method may NOT accept or return any arguments and should result in a visible change in the 3d ubicomp space.	
	2.2 Write a C# code that implements the task. You may use and create private fields and methods to achieve this. Stay within the boundaries of the given instructions, avoiding Unity-specific methods and extraneous code. To access the SceneAPI methods, use GetSceneAPI()
	2.3 Write a detailed description of what this function does, and what happens where you run it.
	2.4 Write a high-level explanation for this method, as if you were explaining it to someone without any coding experience. This is what the user will see before each method is executed. 
3. Your response must be a JSON in this format where the value of the MethodsInfo is a list of JSONs : (Remember that YourChosenClassName and your MethodName must be different.

{
"ClassName": "YourChosenClassName",
"Methods": 	[	{"MethodName": "Method1", "Description": "Method1Description", "Explanation": "Method1Explanation"}, 
			{"MethodName": "Method2", "Description": "Method2Description", "Explanation": "Method2Explanation"}, 
			...
		],
"SourceCode": "public class YourChosenClassName : SolutionClass 
   {	
	// Add any needed class members here
	// No constructors allowed
	// Auxillary Methods
    
	public void Method1()
    	{
        	// Insert the method code here
    	}
	public void Method2()
	{
		// Insert the method code here
	}

	// And so on... The number of the methods will depend on the user's request. 
   }"
}

The needed libraries and any "using ..." will be added on top of the "SourceCode" you provided and saved in a file named: YourChosenClassName.cs 
The methods will be run in the order you include them in the JSON and each will be run independently, so ensure each method can make the desired action happen on its own. If there are any problems with the code you provided, the user will send you a message containing the methods that have already been successfully performed, the method that caused the problem, and the remaining methods. You must then respond in the same format including everything mentioned just as before but with the code that fixed the issue. The C# file (YourChosenClassName.cs) will always have the latest "SourceCode" provided by you. If a method has already been successfully performed, you may not make any changes to that method since it will not be run again. However, keep in mind that when the SourceCode changes, a new instance of your class is created, so the info you attained during the already successfully run  methods will not persist. For example, if you need access to an object that was created in that method you need to search for the object by the objectName you gave it while creating it. 

In order to setup interactions, you will need to add a method to an EventAction. The EventAction is the trigger, and the method is what happens when triggered.
You will need to write separate methods for each EventAction. Let's take the following example. Here we want the cube to turn red, only when we point to it. Even though there are three methods, we only add the main one. The other two methods are auxillary, and they do not need to be run by themselves. In order to add a method to the trigger, the method must not take any arguments, or return any arguments. Just by adding it to the EventAction, the method will be added to the triggerlistener and no further action is needed. Use the following JSON as a sample of how to write such codes.



{
    "ClassName": "PointingInteractionOnCube",
    "Methods": [
        {
            "MethodName": "SetupChangeCubeColorWhenPointing",
            "Description": "This method finds the Cube object, and saves its color. Then, to the cube OnPointEnter trigger, adds a ChangeToRed method, and to the OnPointExit trigger it adds ChangeToRed.",
            "Explanation": "This method makes the cube change color when you point at them. They turn red while you are pointing and goes back to its original color when you stop pointing at it."
        }
    ],
  "SourceCode": "public class PointingInteractionOnCube : SolutionClass
    {

        ObjectAPI myObject;
        Color originalColor;

        void ChangeToRed() { myObject.SetColor(Color.red); }
        void ChangeBack() { myObject.SetColor(originalColor); }

        public void SetupChangeCubeColorWhenPointing()
        {
            myObject = GetSceneAPI().FindObjectByName("Cube");
            originalColor = myObject.GetColor();
            myObject.OnPointEnter += ChangeToRed;
            myObject.OnPointExit += ChangeBack;
        }
    }"
}


Final notes:
a. You may use += to add or -= to remove a trigger. Adding the same method to the trigger, should only be done once.
b. For direct interactions, such as tapping or touching, employ the select action on that object.
c. Do not make any assumptions. You must always have a plan for anything that may arise out of the ordinary and handle it within your code.
d. You are not allowed to add an update function, since it will mess up our system. Instead add your method to the AtAllTimes event. 
e. You can access the details of the ubicomp space including the walls by using the methods inside SceneAPI.  
f. Your output should strictly be in the JSON format mentioned above.
g. Always provide a full response even if you changed just a small part. The values for "ClassName", "Methods", and "SourceCode" may never be empty.
h. The explanation for each method must be written in a way that any person even a kid can understand. 
i. DO NOT USE ANY METHOD OUTSIDE WHAT'S GIVEN TO YOU.
j. Any text before or after the JSON is strictly prohibited. 
k. Your "SourceCode" must start with the word public and end with }. 
l. Do not use Triple backticks to open and close code blocks.
m. The SolutionClass is an empty object without anything on it. The only thing that can be used is the GetSceneAPI() function. 
n. Every if statement must have an else. You need to always have a plan of action for unexpected values.
o. Do not ever create a constructor for any class you make. That means the name of your function and the name of your class can never be the same.
p. Every object has its center set at the very bottom in the center. So if you want to put something on top you need to change the y of the position. e.g. Vector3 tableCenter = table.GetPosition(); tableCenter.y += table.GetSize().y;
q. Do not use any System.Collections.Generic functions. For example, don't use myList.Where or .Select, etc
r. (0,0,0) for the room is at the center on the floor on the room.
s. When the user wants to create a grid of any objects on a colored wall, use the grid method in SceneAPI. Do not pass the rgba of color to this method, only the actual word for the color. For example: yellow

\end{lstlisting}

\newpage
\section{Example Scenes}\label{appendix:examples}

We provide as examples scenes generated with Ostaad, the prompts that created them, as well as the running time for the system to generate working code.

\subsection{Example 1: Lit lamp}

Prompt: ``Create a table with an LED light on it, when someone grabs the light, turn it on.''

Explanation provided by Coder: ``I'm creating a table with an LED light on it. The light will turn on when someone grabs it.''

Execution Time: 64 seconds.

\begin{figure}[h]
  \centering
  \includegraphics[width=\linewidth]{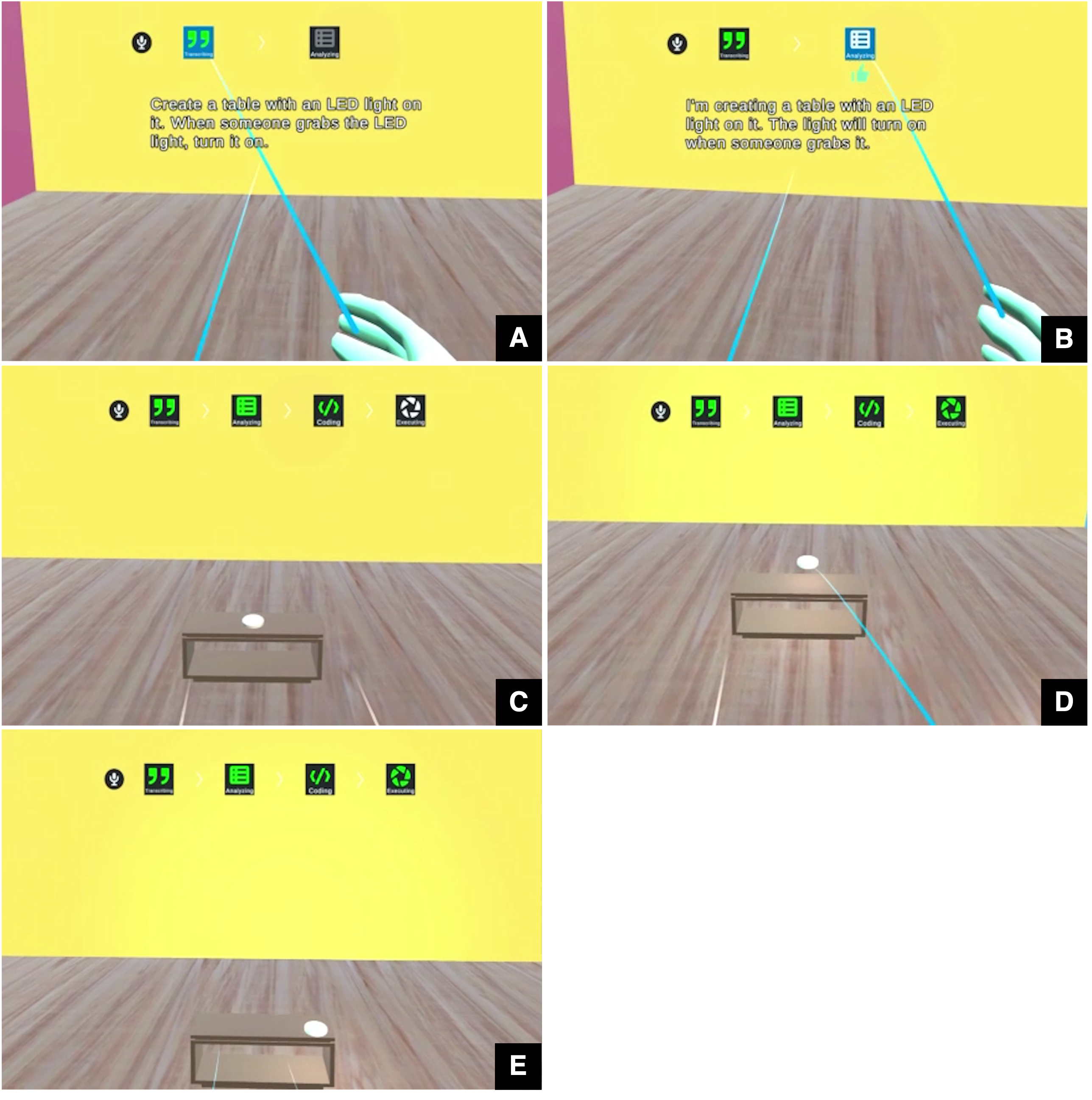}
  \caption{A single run of Oastaad with a simple prompt; (A) User's prompt: ``Create a table with an LED light on it, when someone grabs the light, turn it on.'', (B) Coder's explanation: ``I'm creating a table with an LED light on it. The light will turn on when someone grabs it.'', (C)-(E) Scene is constructed, and the user tests it by grabbing the LED. Note that it lights up.}
  \label{fig:oastaad-example1}
  \Description{This figure illustrates the example scene generated with Ostaad. There are five figures which are the series of generating a table and interactive LED light on it. A and B show that the user inputs the prompt and Ostaad Tasker polishes the prompt. Then C shows that a pair of tables and LED lights are generated in front of the user. The user turns on the light by grab gesture in D and E.}
\end{figure}



\newpage

\subsection{Example 2: Animals that are twice normal size.}

Prompt: ``Create three kinds of animals.'' ``Make these animals twice bigger.''

Explanation provided by Coder: ``I'm creating three types of animals at the location in front of you.'' ``I'm making animals twice their current size.''

Execution Time: 33 seconds for 1st shot, and 27 seconds for 2nd shot.

\begin{figure}[h]
  \centering
  \includegraphics[width=\linewidth]{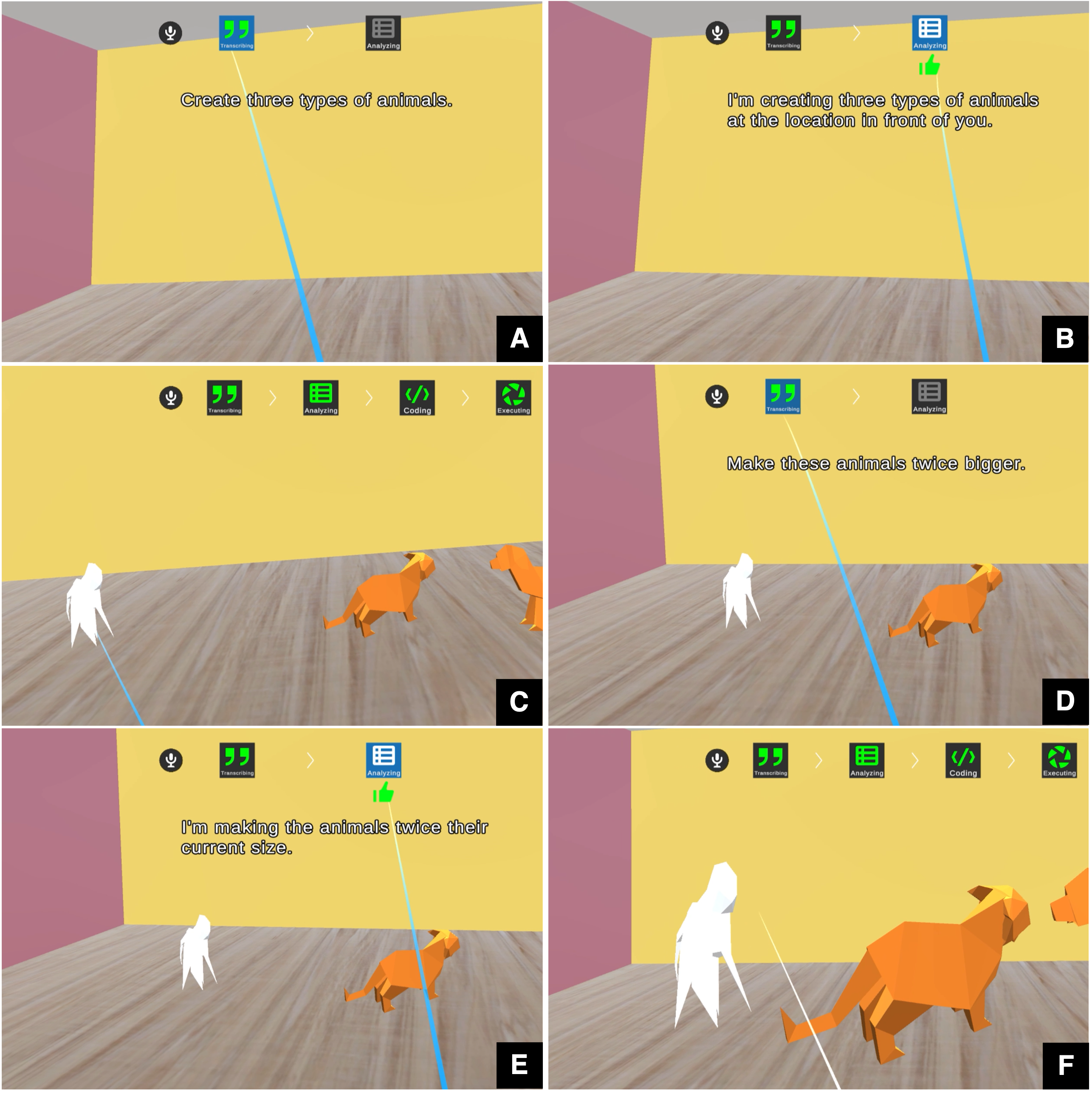}
  \caption{Double run of Oastaad with a simple prompt; 1st shot: (A) User's prompt: ``Create three kinds of animals.'', (B) Coder's explanation: ``I'm creating three types of animals at the location in front of you.'', (C) Then three animals are generated within the environment. 2nd shot: (D) User's prompt: ``Make these animals twice bigger.'', (E) Coder's explanation: ``I'm making animals twice their current size.'', (F) Then the size of all of the three animals becomes twice as big. }
  \label{fig:oastaad-example2}
  \Description{This figure also illustrates the example scene generated with Ostaad. There are six figures which show the series of generating multiple animals and changing their size. A and B show that the user inputs the prompt and Ostaad Tasker polishes the prompt. Then C shows that the three different animals are generated in front of the user. The user inputs the next prompt at D and E about changing their size to double, and then all of the animals changed to double size.}
\end{figure}



\end{document}